\documentclass[sn-basic]{sn-jnl}

\usepackage{graphicx}%
\usepackage{tabularx}  
\usepackage{multirow}%
\usepackage{adjustbox}
\usepackage{chngpage}
\usepackage{amsmath,amssymb,amsfonts}%
\usepackage{amsthm}%
\usepackage{longtable}
\usepackage{mathrsfs}%
\usepackage[title]{appendix}%
\usepackage{xcolor}%
\usepackage{textcomp}%
\usepackage{manyfoot}%
\usepackage{booktabs}%
\usepackage{algorithm}%
\usepackage{algorithmicx}%
\usepackage{algpseudocode}%
\usepackage{listings}%
\usepackage{relsize}
\usepackage{xcolor}

\begin{document}

\title[Article Title]{Relaxation and noise-driven oscillations in a model of mitotic spindle dynamics}


\author*[1]{\fnm{Dionn} \sur{Hargreaves}}\email{dionn.hargreaves@manchester.ac.uk}

\author[1]{\fnm{Sarah} \sur{Woolner}}\email{sarah.woolner@manchester.ac.uk}

\author[2]{\fnm{Oliver} E. \sur{Jensen}}\email{oliver.jensen@manchester.ac.uk}

\affil*[1]{\orgdiv{Faculty of Biology, Medicine and Health}, \orgname{University of Manchester}, \orgaddress{\street{Oxford Road}, \city{Manchester}, \postcode{M13 9PL}, 
\country{UK}}}

\affil[2]{\orgdiv{Department of Mathematics}, \orgname{University of Manchester}, \orgaddress{\street{Oxford Road}, \city{Manchester}, \postcode{M13 9PL}, 
\country{UK}}}


\abstract{During cell division, the mitotic spindle moves dynamically through the cell to position the chromosomes and determine the ultimate spatial position of the two daughter cells. These movements have been attributed to the action of cortical force generators which pull on the astral microtubules to position the spindle, as well as pushing events by these same microtubules against the cell cortex and {\color{black}{plasma}} membrane. Attachment and detachment of cortical force generators working antagonistically against centring forces of microtubules have been modelled previously (Grill \hbox{et al.} 2005, \textit{Phys. Rev. Lett.} \textbf{94}:108104) via stochastic simulations and {\color{black}mean-field} Fokker--Planck equations {\color{black}(describing random motion of force generators)} to predict oscillations of a spindle pole in one spatial dimension. 
Using systematic asymptotic methods, we reduce the Fokker--Planck system to a set of ordinary differential equations (ODEs), consistent with a set proposed by Grill et al., which {\color{black}can} provide accurate predictions of the conditions for the Fokker--Planck system to exhibit oscillations.
In the limit of small restoring forces, we derive an algebraic prediction of the 
amplitude of spindle-pole oscillations and demonstrate the relaxation structure of nonlinear oscillations.  We also show how noise-induced oscillations can arise in stochastic simulations for conditions in which the {\color{black}mean-field} Fokker--Planck system predicts stability, but for which the period can be estimated directly by the ODE model {\color{black} and the amplitude by a related stochastic differential equation that incorporates random binding kinetics.}}




\maketitle

\section{Introduction}\label{sec1}

\begin{figure}
\centering
\includegraphics[width=0.9\textwidth]{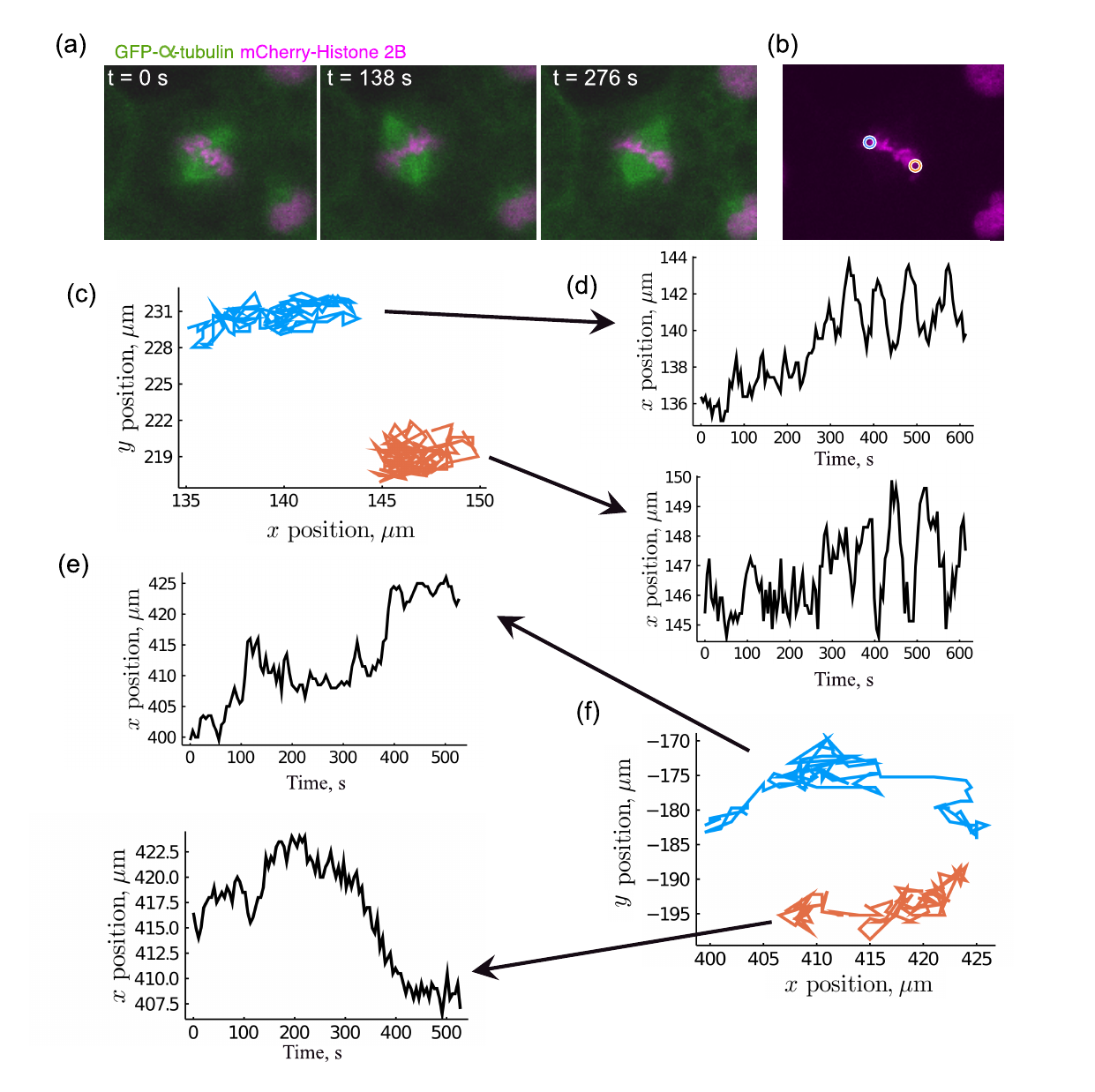}
\caption{\textbf{The mitotic spindle and metaphase plate oscillate during the metaphase stage of mitotic cell division.} (a) Time-lapse images of a mitotic spindle (GFP-$\alpha$-tubulin, green) and metaphase plate (mCherry-Histone 2B, magenta) during metaphase of a cell dividing in a $\emph{Xenopus laevis}$ embryo, at stage 10-11. The metaphase plate lies perpendicular to the fusiform shape of the mitotic spindle. (b) Blue and orange circles indicate the measured termini of the metaphase plate. (c, f) Tracked positions of metaphase-plate termini over a full course of metaphase, for two cells in an excised \emph{Xenopus} animal cap at stage 10-11. (d) The $x$-components of the termini tracked in (c), showing oscillatory motion as a function of time. (e) The $x$-components of the termini of the metaphase plate tracked in (f), showing non-oscillatory motion as function of time. Arrows indicate the relevant measured terminus. Data from \cite{Hargreaves2023mitotic}, obtained using methods described in Appendix~\ref{sec:A3}.}
\label{fig:example}
\end{figure}

Embryos develop, on the most basic level, as a result of one cell dividing into two cells. In a tissue, the orientation of cell division is an important factor in determining either the outcome for the daughter cells (e.g. cell fate due to distribution of intracellular components or the daughter cell local environment) or the tissue as a whole (e.g. building tissue and organ architecture by tissue stratification or spreading) \citep{bergstralh2014spindle, morin2011mitotic}. Cell division orientation is determined by the mitotic spindle, the large microtubule-based structure which forms in the cell and segregates genetic material into two discrete daughter cells \citep{karsenti2001mitotic, mitchison2001mitosis}. Prior to anaphase, where the chromosomes are pulled apart to opposite ends of the cell, the mitotic spindle is positioned translationally and rotationally (Fig.~\ref{fig:example}).

Key to spindle positioning is the pushing and pulling of astral microtubules which reach between the spindle pole and the cell cortex \citep{dogterom2005force,burakov2003centrosome, zhu2010finding,pecreaux2006spindle,howard2006elastic, okumura2018dynein, bosveld2016epithelial, pecreaux2016mitotic}. Pulling is mediated at the cell cortex through interactions between astral microtubules and the motor protein dynein, which is anchored at the plasma membrane through its association with a tripartite complex consisting of NuMA, LGN and G$\alpha$i \citep{okumura2018dynein, bosveld2016epithelial, pecreaux2016mitotic}. Mathematical models have been used to investigate how astral microtubule pushing and pulling can drive spindle positioning. 
For example, minimising the calculated torque, created by pulling forces along the microtubule length \citep{minc2011influence} or by concentrated populations of dynein-associated proteins at the cell periphery \citep{thery2007experimental,bosveld2016epithelial}, has been shown to predict the cell division orientation in sea urchin zygotes \citep{minc2011influence}, micropattern-adhered HeLa cells \citep{thery2007experimental} and the \emph{Drosophila} pupal notum epithelium \citep{bosveld2016epithelial}. These models highlight the importance of cell geometry \citep{minc2011influence} and the localisation of dynein \citep{thery2007experimental,bosveld2016epithelial} in spindle orientation and consequently cell division orientation. In a different approach, \cite{li2017geometric} used a stochastic model to describe the interactions of microtubules with chromosomes, motor proteins and boundaries to create self-assembled spindles within cells. 
This model was adapted to investigate spindle orientation: microtubules and chromosomes self-assemble into a mitotic spindle and orient within the simulated cells as a result of a combination of microtubule pushing forces and dynein-mediated pulling forces at both the cortex and within the cytoplasmic domain \citep{li2019cell}, resulting in spindles which align with sites of localised dynein 
similarly to what has been shown in simpler models \citep{thery2007experimental,bosveld2016epithelial}. Interestingly, the simulated spindles were shown to form already in line with their final division axis, with no notable movements of the spindle once assembled. 

However, the mitotic spindle has been observed to approach its final destination less directly after its assembly. During the first division of the \emph{C. elegans} fertilised egg, the posterior spindle pole undergoes a defined oscillation as the mitotic spindle is asymmetrically positioned in the cell to produce two daughter cells of unequal sizes  \citep{pecreaux2006spindle,pecreaux2016mitotic}. Similarly, the rotational movements of the spindle in \emph{Xenopus} epithelial tissue have been shown to be dynamic, culminating in oscillations of the spindle angle immediately prior to anaphase \citep{larson2017automated}. In the developing airway epithelium of mice, a subset of cells have been identified which continuously change their mitotic spindle angle throughout metaphase \citep{tang2018mechanical}.  {\color{black}Different strains of nematodes related to \emph{C.~elegans} show robust spindle oscillations, albeit with inter- and intra-specific variations in dynamical features\citep{valfort2018}.}

In the context of highly coordinated spindle movements, such as the oscillation of the \emph{C.~elegans} zygote posterior pole, contributions from both microtubule pushing and dynein-mediated pulling have been shown mathematically to produce the observed oscillatory dynamics (\cite{grill2005theory, pecreaux2006spindle}{\color{black}{; reviewed in \cite{beta2017intracellular}}}). This mathematical model describes changes in the position of the mitotic spindle pole in 1D as a result of pulling by cortical force generators, combined with microtubule-based restoring forces \citep{grill2005theory}. The relative simplicity of this model compared with that used by \cite{li2017geometric} lends itself more readily to an investigation of the primary parameters giving rise to dynamic movements and, crucially, replicates the oscillations seen in \emph{C. elegans}.  {\color{black}{\cite{wu2024laser} also describe the \textit{C. elegans} zygote posterior pole oscillation, using a 2D model omitting pushing from microtubules. This omission is congruent with their analysis of subcellular fluid flows, which suggests that cortical pulling forces dominate to drive movement. The region of the cortex subtending the microtubule array as the spindle pole approaches and recedes from the cortex is highlighted as a key factor for creating oscillations in the pole position. As a result, pushing forces are unnecessary to create a reversal of the spindle pole velocity, as the direction of pulling by force generators is re-distributed across angles away from the cortex upon approach of the spindle pole \citep{wu2024laser}. Strikingly, the resulting oscillations are nonlinear, in contrast with those demonstrated by \cite{grill2005theory}, though the significance of these nonlinear oscillations has yet to be explored.}} The {\color{black}{correct}} balance of microtubule pushing and dynein-mediated pulling forces are likely the drivers for producing spindle movements and subsequently for determining the division orientation.

Other cells demonstrate spindle movements which are more complex. By mathematical amplification of pulling forces on the mitotic spindle from discrete cortical locations, rotational spindle dynamics 
have been simulated to match those observed in HeLa cells \citep{corrigan2015modeling}. Stochastic switching between active and non-active cortical cues simulates noisy rotation toward the long axis of the cell as defined by anisotropy in the placement of the cortical locations \citep{corrigan2015modeling}, highlighting both the importance of cortical cue elements in spindle orientation and the possibility of stochasticity in creating dynamic movements of the spindle. Stochastic processes can result in behaviours which are not captured by deterministic models due to processes such as stochastic resonance \citep{erban2020}. Indeed, the addition of noise inherent to biological systems \citep{tsimring2014noise} should not be discarded in considerations of dynamic behaviour.

In \emph{Xenopus} embryo epithelial tissue, mitotic spindles have been shown to undergo both a net rotation towards the final division axis and a stereotypical oscillation prior to anaphase onset \citep{larson2017automated} (see Online Resource 1). Fig.~\ref{fig:example}b-d illustrates such oscillations in \emph{Xenopus} animal cap epithelial tissue, obtained by tracking the movement of the metaphase plate, with which mitotic spindle movements are highly correlated.   
Oscillations are noisy with a nonlinear structure suggestive of relaxation oscillations (with rapid reversals of direction, Fig.~\ref{fig:example}(d)), a feature not reportedly observed in the \emph{C. elegans} spindle oscillation \citep{pecreaux2006spindle}.
The factors which affect the structure of oscillations in the mitotic spindle, specifically the nonlinear structure identified in Fig.~\ref{fig:example}(d), have not yet been fully described, {\color{black}{although spindle movements driven by cortical force generators in the absence of microtubule pushing forces appear to create nonlinear oscillations  \citep{wu2024laser}, suggesting that relaxation oscillations may arise in pull-dominated systems.}} Furthermore, spindles which do not oscillate are also present {\color{black}{within the same tissue}} (Fig.~\ref{fig:example}(e,f)),  in contrast to the defined and characteristic spindle behaviour of the \emph{C. elegans} zygote \citep{pecreaux2006spindle,pecreaux2016mitotic}. It is unclear how more complex tissue environments, such as is found in the \emph{Xenopus} epithelium, may affect the ability of mitotic spindles to oscillate, or the non-linearity of the oscillatory spindle movements, motivating the present study of spindle dynamics.

In this paper, we revisit the mathematical model presented by \cite{grill2005theory}, investigating factors which promote relaxation and noise-driven oscillations. The model is outlined in Section~\ref{sec:model} and stochastic simulations are presented in Section~\ref{Gillespie}.  Representations of solutions using {\color{black}mean-field} Fokker--Planck equations {\color{black} which account for noise in the random walking of force generators, but which incorporate a deterministic representation of binding kinetics,} are given in Section~\ref{sec:FP}; these are then reduced to a set of ordinary differential equations (ODEs) using systematic asymptotic analysis in Section~\ref{sec:odes} {\color{black} assuming slow binding kinetics}.  The ODEs turn out to represent a special case of the (less formally derived) ODEs proposed by \cite{grill2005theory}. A stability analysis gives predictions for the period of oscillation at the onset of neutral oscillations, as well as the position of the neutral oscillation boundary in parameter space, in agreement with Fokker--Planck predictions. 
Further asymptotic reduction of the ODE model {\color{black}{in the limit of small pushing forces}} (Section~\ref{sec:relax}) yields a single algebraic equation which describes the structure of nonlinear relaxation oscillations.  We provide {\color{black}numerical} evidence that smaller-amplitude irregular oscillations, characteristic of observations (Fig.~\ref{fig:example}), can be induced by noise {\color{black}associated with random binding kinetics} through stochastic resonance. {\color{black}  This is supported by analysis in Section~\ref{sec:noisosc} of a stochastic differential equation, derived from the ODE model, that seeks to estimate the amplitude and spectrum of the noise-induced oscillations.}

\section{The model of spindle pole dynamics}
\label{sec:model}

\begin{figure}
    \centering
    \includegraphics[width=0.9\textwidth]{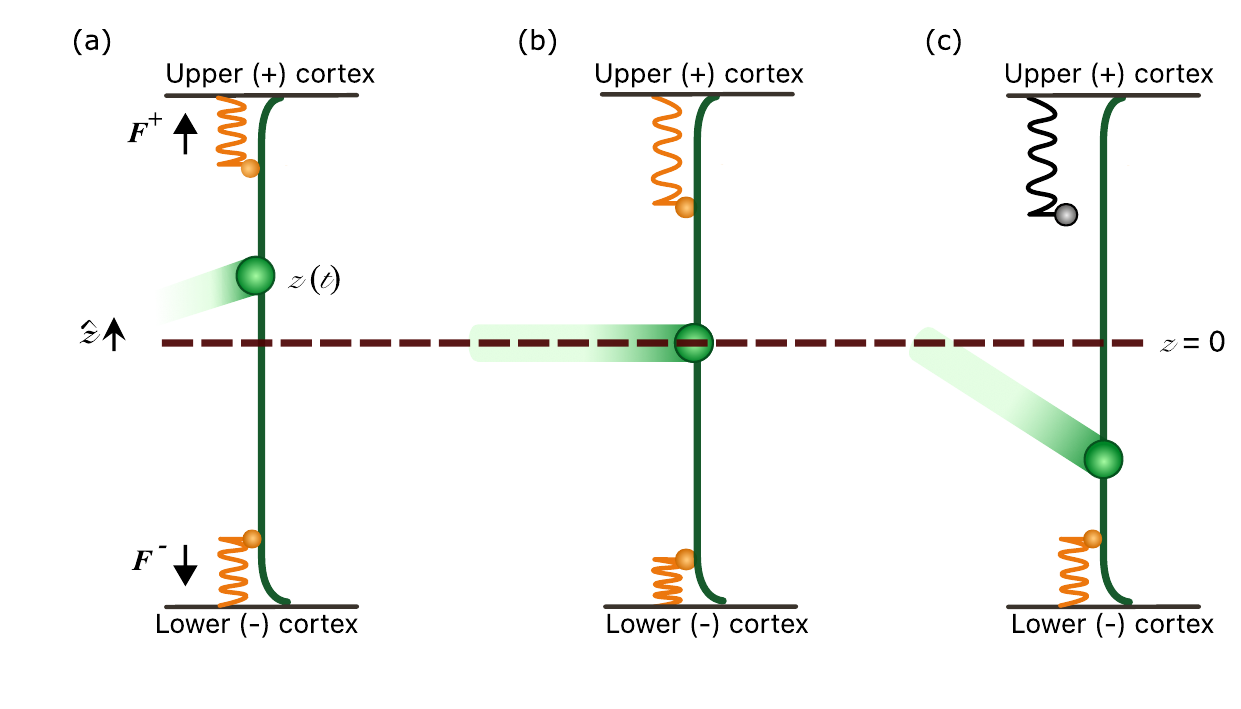}
    \caption{\textbf{Diagram of a spindle pole in three states.} (a) The spindle pole (green) lies between the upper and lower cortex, displaced a distance ${z}({t})$ from the mid-point. Force generators (orange) at each cortex comprise a motor protein head and an elastic linker which produce pulling forces ${F}^\pm$. (b) Movement of the spindle pole affects the linker extensions of the motor proteins: movement away from the upper cortex lengthens the linkers of the upper force generators while compressing the linkers of the lower force generators. (c) Force generators with more extended linkers have an increased unbinding rate. Unbound generators cannot produce a pulling force (indicated by a grey force generator).}
    \label{fig:PDEdiag}
\end{figure}

In the 1D model of spindle pole dynamics proposed by \cite{grill2005theory}, a spindle pole at position $\bar{z}\left(\bar{t}\right)$ at a time $\bar{t}$ moves along an axis $\mathbf{\hat{z}}$ spanning a cell according to 
\begin{equation}\label{eq:polepos}
    \bar{\xi}\frac{\textrm{d}\bar{z}}{\textrm{d}\bar{t}} + k_{\textrm{MT}}\bar{z}\left(\bar{t}\right) = \bar{F}^+-\bar{F}^-.
\end{equation}
The parameter $\bar{\xi}$ models viscous drag on the pole from the cytoplasm.  The stiffness parameter $k_{\mathrm{MT}}$ represents a restoring force towards the cell midplane, arising from dynamic instability and bending of astral microtubules that emanate from the spindle pole and extend to the cell cortex \citep{grill2005theory, pecreaux2006spindle, pecreaux2016mitotic, howard2006elastic, rubinstein2009elasticity}. 
The spindle is pulled towards either side of the cell under fluctuating forces $\bar{F}^\pm(\bar{t})$.  Pulling arises from individual force generators which lie at the cell cortex and bind to astral microtubules. For simplicity, the model considers two opposing populations of force generators which sit in an `upper' and `lower' cortex, labelled $\pm$ hereafter. The force generators comprise a motor protein head connected to the cortex via an elastic linker of stiffness $k_{\textrm{g}}$ (Fig.~\ref{fig:PDEdiag}). The motor protein head can be considered to be dynein, which binds to and walks along the microtubules towards the spindle pole. The two populations of $N$ force generators are assumed to exert pulling forces toward their respective cortex with a magnitude 
\begin{equation} \label{eq:popforcing}
\bar{F}^{\pm}(\bar{t})=k_{\textrm{g}}{\textstyle{\sum}}_{n=1}^{N}\bar{y}^{\left(n\right)\pm}_{\textrm{b}}(\bar{t}),
\end{equation}
where $\bar{y}^{\left(n\right)\pm}_{\textrm{b}}(\bar{t})$ is the extension of the elastic linker of bound (subscript b) force generator $n$. In (\ref{eq:popforcing}) it is assumed that the pulling force due to an individual linker is proportional to its length. Unbound (subscript u) force generators of length $\bar{y}^{\left(n\right)\pm}_{\textrm{u}}$ (Fig.~\ref{fig:PDEdiag}(c)) are not connected to the spindle pole and are unable to provide any forcing.  The superscript ${(n)}$ in (\ref{eq:popforcing}) is used to label the $n_{\mathrm{b}}^\pm$ linkers that, in any short interval $(\bar{t}, \bar{t}+\delta \bar{t})$, are bound to a microtubule, where $0\leq n_{\mathrm{b}}^\pm\leq N$.

The forcing in (\ref{eq:polepos}) fluctuates because the linkers bind and unbind randomly. The movements of the spindle pole are tightly coupled to the individual extension lengths of the linkers via  (\ref{eq:popforcing}), and by the fact that spindle motion influences the length of bound linkers. The motor protein heads have walking velocities given by
\begin{subequations}
\label{eq:vels}
\begin{equation}\label{eq:velocity}
    {\bar{v}^{\left(n\right)\pm}_\textrm{b}}=v_0\left(1-\frac{k_{\textrm{g}}\bar{y}^{\left(n\right)\pm}_{\textrm{b}} }{f_0}\right)\mp\frac{\textrm{d}\bar{z}}{\textrm{d}\bar{t}}.
\end{equation}
Here $f_0$ is the stall force of a force generator, \hbox{i.e.} the force required to bring the motor protein head to rest relative to the spindle pole; it is assumed that the unloaded walking velocity $v_0$ is reduced in proportion to $k_{\textrm{g}}\bar{y}^{(n)\pm}_{\textrm{b}}/f_0$, the tensile force acting upon the motor protein head by the elastic linker scaled relative to $f_0$.  Equivalently, $y_0\equiv f_0/k_g$ is the extension of a linker at which it stalls.  In (\ref{eq:velocity}), the spindle pole velocity term $\mp{\textrm{d}\bar{z}}/{\textrm{d}\bar{t}}$ arises due to the force generator being connected to the moving spindle pole via the microtubules.  Thus, as the spindle pole moves towards a bound force generator it will compress the elastic linker by pushing on the bound motor, reducing its relative walking velocity (Fig.~\ref{fig:PDEdiag}). 
After a linker detaches from a microtubule, becoming unbound, it contracts with velocity 
\begin{equation}\label{eq:velocityunb}
    {\bar{v}^{\left(n\right)\pm}_\textrm{u}}=-(k_{\textrm{g}}/\bar{\xi}_g)\bar{y}^{\left(n\right)\pm}_{\textrm{u}},
\end{equation}
\end{subequations}
where $\bar{\xi}_g$ is a drag coefficient of unbound dynein.  Using the largest proteins in the force generator complex (dynein, length approximately 50~nm \citep{trokter2012reconstitution},  and NuMA, length approximately 210~nm \citep{compton1993numa}), a force generator has a Stokes radius of order $10^{-1}$ smaller than the spindle pole, so that $\bar{\xi}_{\textrm{g}}\approx \bar{\xi}\times10^{-1}$.  The superscript ${(n)}$ in (\ref{eq:velocityunb}) is used to label the $n_{\mathrm{u}}^\pm$ linkers that, in the short interval $(\bar{t}, \bar{t}+\delta \bar{t})$, are unbound, where $0\leq n_{\mathrm{u}}^\pm\leq N$ and $n_{\mathrm{u}}^\pm+n_{\mathrm{b}}^\pm=N$

The model is closed by relating linker lengths to linker velocities, incorporating noise in the linker dynamics through effective diffusion coefficients $\bar{D}_{\mathrm{u}}$ and $\bar{D}_{\mathrm{b}}$, and by modelling the transitions between bound and unbound states as random events taking place at rates $\bar{\omega}_{\mathrm{on}}$ and $\bar{\omega}_0 \exp[\bar{\gamma}\bar{y}_{\mathrm{b}}^{(n)\pm}]$ respectively.  Here $\bar{\gamma}$ parametrizes the slip-like manner in which dynein detaches from microtubules under loading \citep{ezber2020dynein}.  Estimated values of dimensional parameters are summarized in Table~\ref{table:params}.


\newcolumntype{L}{>{\centering\arraybackslash}m{5cm}}
\begin{table}
\centering
\begin{tabular}[width=0.03\textwidth]{|L|c|c|c|}
\hline 
Description & Parameter & Value & Reference \\  [0.1cm] \hline 
& & & \\

Drag coefficient & $\bar{\xi}$ & $10^{-6}$~Nsm$^{-1}$& 1  \\  [0.1cm]

Microtubule stiffness & $k_{\textrm{MT}}$ & $4\times10^{-6}$~Nm$^{-1}$ & 1, 2  \\ [0.25cm]

Elastic linker stiffness & $k_{\textrm{g}}$  & $8\times10^{-5}$~Nm$^{-1}$  & 1  \\ [0.25cm]

Stall force & $f_0$ & $3\times10^{-12}$~N &  1, 3, 4  \\ [0.25cm]

Spontaneous velocity of force generators & $v_0$ & $1.8\times10^{-6}$~ms$^{-1}$& 1, 5 \\ [0.25cm]

Stall rate
& $
v_0/y_0$ & $50$~s$^{-1}$ & 1  
\\ [0.25cm]

Retraction rate of unbound generators & $
k_g/\bar{\xi}_g$  & $10^3$~s$^{-1}$ & (1, 6, 7) \\  [0.3cm]

Sensitivity of unbinding to linker extension & $\bar{\gamma}$ & $5.6\times10^7$~m$^{-1}$ & (1) \\ [0.25cm]

Diffusion coefficient of bound generators & $\bar{D}_{\textrm{b}}$ & $5\times10^{-15}$~m$^2$s$^{-1}$ & 1 \\ [0.25cm]

Diffusion coefficient of unbound generators & $\bar{D}_{\textrm{u}}$ & $5\times10^{-14}$~m$^2$s$^{-1}$ & (1) \\[0.25cm]

Number of force generators per cortex & $N$ & - & \\ [0.25cm]

Maximum linker extension & $\bar{y}_{\textrm{max}}$ & $2.16\times10^{-7}$~m & (1) \\ [0.25cm]

Microtubule-generator binding rate & $\bar{\omega}_{\textrm{on}}$ & $0.15$~s$^{-1}$ &\\ [0.25cm]

Microtubule-generator unbinding rate coefficient & $\bar{\omega}_{0}$ & $0.05$~s$^{-1}$ & 1 \\ [0.25cm]

\hline
\end{tabular} 
\caption{Parameter values and descriptions. References in parenthesis contain information which was used in order to derive the parameter value. References: 1 \cite{grill2005theory}; 2 \cite{rubinstein2009elasticity}; 3 \cite{belyy2014cytoplasmic}; 4 \cite{ezber2020dynein}; 5 \cite{milo2015cell}; 6 \cite{harborth1995epitope}; 7 \cite{ trokter2012reconstitution} }\label{table:params}

\end{table}

\begin{table}
\centering
\begin{tabular}[width=\textwidth]{|c|c|c|}
\hline
Parameter & Components  &  Baseline value  \\ \hline \hline
&  & \\
 ${\xi}$ & 
 $\bar{\xi}v_0/f_0$
 & 0.625 \\[0.15cm]
 & &\\
 $K$&$\displaystyle{{k_{\textrm{MT}}}/{k_{\textrm{g}}}}$ & 0.05 \\[0.15cm]
 & &  \\
 ${\omega}_{\textrm{on}} $&
 $\bar{\omega}_{\mathrm{on}} y_0/v_0$
 & 0.003 \\[0.15cm]
 & &  \\
${\omega}_{0}$&
$\bar{\omega}_0 y_0/v_0$
& 0.001 \\[0.15cm]
& & \\
${y}_{\textrm{max}}$ & 
$\bar{y}_{\max}/y_0$
& 6\\[0.15cm]
& &\\
$\gamma$ & 
$\bar{\gamma} y_0$
& 2 \\[0.15cm]
& &\\
 $D_{\textrm{b}} $&
$\bar{D}_{\mathrm{b}}/(y_0v_0)$
 & 0.08 \\[0.15cm]
 & &\\
$D_{\textrm{u}} $&
$k_{\mathrm{b}}T/(v_0 f_0)$
& 0.04 \\[0.15cm]
& & \\
 $\Gamma$& 
$f_0/(\bar{\xi}_g v_0)$
 & 20 \\[0.15cm]\hline
\end{tabular} 
\caption{Nondimensional parameters are given in terms of dimensional parameters. 
Baseline values are used in figures below, except where indicated.  $k_{\mathrm{b}}T$ is the unit of thermal energy.} \label{table:nondims}
\end{table}

Scaling lengths on the stall length $y_0$ and time on the stall time $y_0/v_0$ (so that $\bar{z}=y_0 {z}$, $\bar{t}=(y_0/v_0){t}$, etc.), the force balance on the spindle pole (\ref{eq:polepos}, \ref{eq:popforcing}) becomes, in dimensionless form,
\begin{equation}\label{eq:polepos_nondim}
    {\xi}\frac{\textrm{d}{z}}{\textrm{d}{t}} = - K{z}\left({t}\right) + \left(\textstyle{\sum}_{n'=1}^{N}{y}^{(n')+}_{\textrm{b}}\left({t}\right)-\textstyle{\sum}_{n=1}^N{y}^{(n)-}_{\textrm{b}}\left({t}\right)\right).
\end{equation}
${\xi}=\bar{\xi}v_0/(k_{\textrm{g}}y_0)$ and $K=k_{\textrm{MT}}/k_{\textrm{g}}$ are dimensionless drag and stiffness parameters respectively. The velocities of the bound and unbound generators (\ref{eq:velocity}) become
\begin{equation}\label{eq:velocity2_nondim}
    {{v}^{\left(n\right)\pm}_\textrm{b}}=1- {y}^{\left(n\right)\pm}_{\textrm{b}}\mp\frac{\textrm{d}{z}}{\textrm{d}{t}}, \quad
    {{v}^{\left(n\right)\pm}_\textrm{u}}=- \Gamma{y}^{\left(n\right)\pm}_{\textrm{u}},
\end{equation}
where $\Gamma=f_0/(\bar{\xi}_g v_0)$.  This parameter measures 
$k_g/\bar{\xi}_g$, the retraction rate of unbound linkers, relative to $v_0/y_0$, the stall rate.  
Dimensionless counterparts of the stochastic parameters are diffusion coefficients ${D_{\mathrm{b}}}$ and ${D_{\mathrm{u}}}$ {\color{black}describing the mobility of} bound and unbound linkers respectively, and transition rates ${\omega}_{\mathrm{on}}$ and ${\omega}_0 e^{\gamma {y}_{\mathrm{b}}^{(n)\pm}}$ respectively.  Dimensionless parameters are summarised in Table~\ref{table:nondims}.

Over any short interval, the populations of bound and unbound linkers have a distribution of lengths.  Average extensions are defined by 
\begin{equation}
\left\langle{y}^\pm_{\textrm{b(u)}}\right\rangle=\frac{{\sum}_{n=1}^{N}{y}^{(n)\pm}_{\textrm{b(u)}}}{n^\pm_{\textrm{b(u)}}}.
\label{eq:avext}
\end{equation}


\section{Stochastic simulations}
\label{Gillespie}

To capture the discrete interactions between a small number of force generators and the spindle pole, we 
discretize ${y}^{\left(n\right)\pm}_{\textrm{b}}$ in increments of $\Delta {y}$ and use a Gillespie algorithm to model the stochastic extensions and retractions of bound and unbound linkers 
and the stochastic transitions of the binding state of the force generators, as they bind and unbind from microtubules.  As explained in Appendix~\ref{secA1}, the extension and retraction of the force generators are treated as $2N$ biased random walks with drift ${v}_{\textrm{b(u)}}^\pm$, diffusion ${D_{\mathrm{b(u)}}}$ and state change (between bound and unbound states), all coupled to displacement of the spindle.

\begin{figure}
\includegraphics[width=\textwidth]{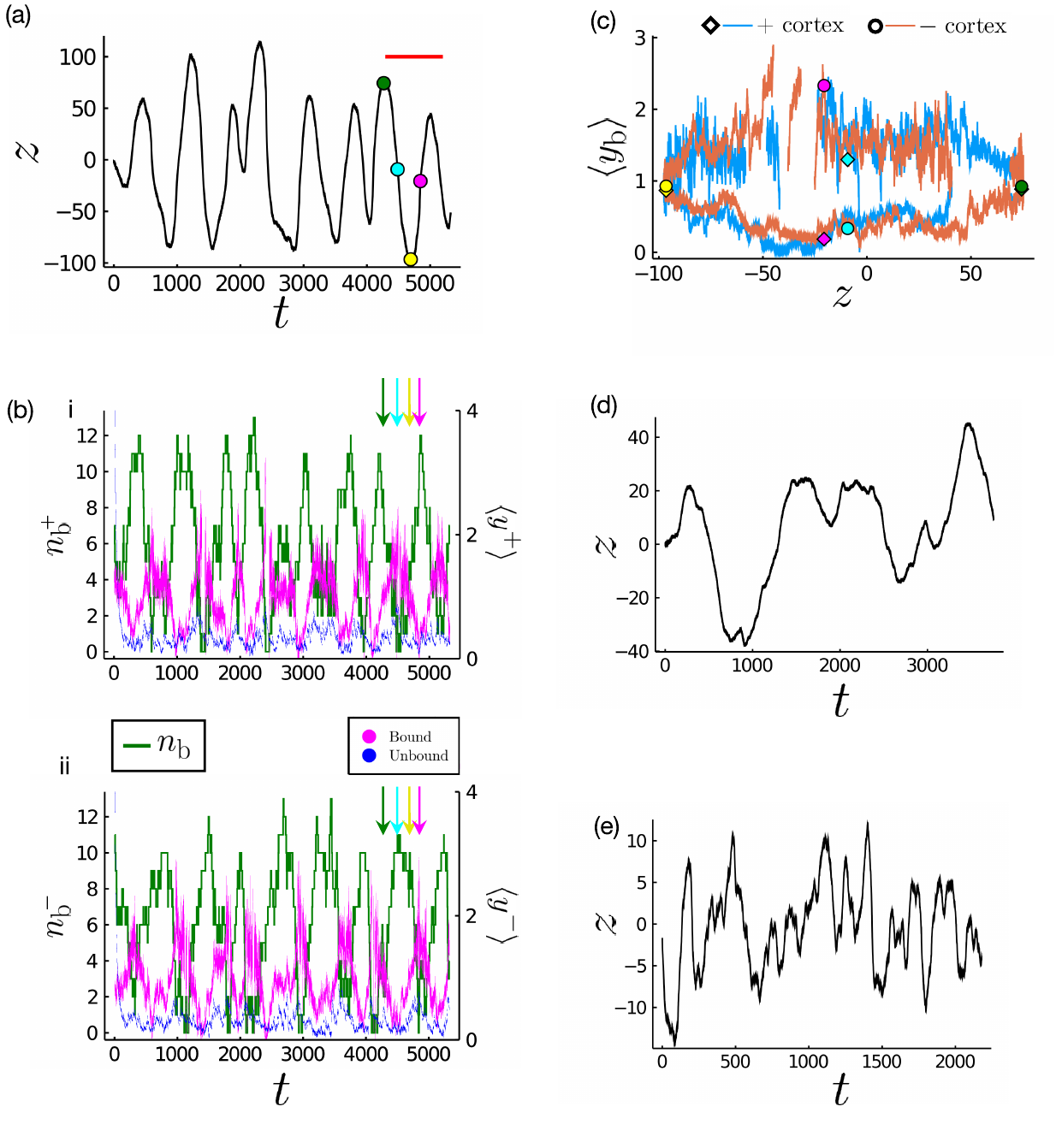}
\caption{\textbf{Stochastic simulations can predict spontaneous oscillations of the spindle pole position.} (a) Evolution of the non-dimensionalised spindle pole position through time. Dots correspond to moments in the cycle of interest and correspond colour-wise with the dots and diamonds plotted in (c). {\color{black}For later reference,} the red bar identifies the oscillation period predicted by (\ref{eq:fullperiod}) below. (b) The number of bound force generators (green) in the i) upper (+) and ii) lower (-) cortex (left $y$-axis) through time. The average extensions of the bound (magenta) and unbound (blue) force generators in the i) upper (+) and ii) lower (-) cortex are also shown (right $y$-axis). Coloured arrows correspond temporally to coloured symbols in (a).  (c) Average extension of the bound generators in the upper (blue) and lower (orange) cortices as a function of pole position. Parameters in (a,b,c) are as in Table~\ref{table:nondims} with $N=15$. (d) A simulation when the unbinding of the force generator is no longer tension-sensitive, with $\gamma=0$.  (e) A simulation when the restoring force is increased by a factor of 100 to $K=5$.}
\label{fig:stochastic_ex}
\end{figure}

Fig.~\ref{fig:stochastic_ex} presents a simulation displaying the emergence of spontaneous oscillations of the spindle pole, using the parameters shown in Table \ref{table:params}, with $N=15$ linkers at either cortex.  The spindle location (Fig.~\ref{fig:stochastic_ex}(a)), numbers of bound and unbound force generators $n^\pm_{\textrm{b(u)}}$ (Fig.~\ref{fig:stochastic_ex}(b)) and average extensions  
$\langle{y}^\pm_{\textrm{b(u)}}\rangle$ (\ref{eq:avext}) (Fig.~\ref{fig:stochastic_ex}(b,c)) show noisy but oscillatory dynamics.
The average extensions of the bound force generators $\langle{y}^+_{\textrm{b}}\rangle$ and $\langle{y}^-_{\textrm{b}}\rangle$ (\ref{eq:avext}) oscillate in anti-phase to one another (Fig.~\ref{fig:stochastic_ex}(b)).  The average extension of unbound force generators $\langle{y}^\pm_{\textrm{u}}\rangle$ remains close to $0$ following initial transients (Fig.~\ref{fig:stochastic_ex}(b)). This can be explained by considering the movement of the spindle pole through one cycle of oscillation (Fig.~\ref{fig:stochastic_ex}(a)) and $\langle{y}^\pm_{\textrm{b}}\rangle\left({z}\right)$ (Fig.~\ref{fig:stochastic_ex}(c)), discussed further below.  Apparent gaps in the $\langle{y}^\pm_{\textrm{b}}\rangle$ plots (Fig.~\ref{fig:stochastic_ex}(c)) occur where there are no bound generators from which to extract an average (where $n_{\textrm{b}}$ = 0 in Fig.~\ref{fig:stochastic_ex}(b)).

Consider the following phases of movement identified by coloured symbols in Figs~\ref{fig:stochastic_ex}(a,c).
\begin{enumerate}
    \item {\textbf{Spindle moving away from the upper cortex (green to cyan)}}. At the peak of the spindle pole oscillation, movement of the spindle is dominated by the microtubule restoring force. The bound generators are extended equally in the upper and lower cortices ($\langle{y}^+_{\textrm{b}}\rangle \sim \langle{y}^-_{\textrm{b}}\rangle$ at the green timepoint (Fig.~\ref{fig:stochastic_ex}(c)) )though there are a greater number bound in the upper cortex rather than the lower ($n_{\textrm{b}}^+ > n_{\textrm{b}}^-$, comparison in Fig.~\ref{fig:stochastic_ex}(bi) vs (bii)). The restoring force ($-K{z}$) is greater than the net upward pulling force provided by this unbalanced population ratio. As the spindle pole moves towards ${z}=0$, this restoring force decreases while the increasing spindle pole velocity results in a net compression of the elastic linkers on the lower cortex, due to a switch in the sign of ${v}^-_{\textrm{b}}\left(\langle{y}^-_{\textrm{b}}\rangle\right)$. Additionally, the spindle pole velocity increases the relative velocity of the force generators in the upper cortex, resulting in an extension of the elastic linkers at the upper cortex (Figs~\ref{fig:stochastic_ex}(c)), and shortening of the linkers at the lower cortex. Due to the tension-sensitive unbinding rate ${\omega}_0e^{\gamma{y}^+_{\textrm{b}}}$, this results in a gradual decrease in the number of upper bound force generators as $\langle{\omega}_0e^{\gamma{y}^+_{\textrm{b}}}\rangle$ increases in value, while the number of bound force generators in the lower cortex increases due to a constant binding rate and a decreased unbinding rate (Fig.~\ref{fig:stochastic_ex}(b)). 
    \item {\textbf{Spindle moving through the centre of its oscillating range, toward the lower cortex (cyan to yellow)}}. As the spindle moves through ${z}=0$ the restoring force steadily increases from $0$ to $-K{z}$. This slows the movement of the spindle such that the velocity of the force generators in the lower cortex may become positive ${v}^-_{\textrm{b}}\left(\langle{y}^-_{\textrm{b}}\rangle\right)>0$ which allows these elastic linkers to extend (Fig.~\ref{fig:stochastic_ex}(c)), decreasing the relative velocity of the remaining upper force generators, the average extension of which is also reduced due to the unbinding of those with larger extensions and binding of force generators with reduced extensions (Fig.~\ref{fig:stochastic_ex}(c)). The number of bound generators in the lower cortex also begins to decline as they extend due to the increased unbinding rate (Fig.~\ref{fig:stochastic_ex}(bii)).
    \item {\textbf{Spindle moving away from the lower cortex (yellow to magenta)}}. This phase replicates the first phase, but with the behaviours of upper and lower cortex reversed. The motion away from the cortex due to the restoring force results in a compression of the upper elastic linkers and an extension of the lower elastic linkers (Fig.~\ref{fig:stochastic_ex}(c)), and a corresponding decrease in the absolute number of bound force generators in the lower cortex as opposed to the increased binding observed in the upper cortex (Fig.~\ref{fig:stochastic_ex}(b)).
\end{enumerate}

The closed loops in $(\langle{y}^\pm_{\textrm{b}}\rangle, {z})$ space are traced anti-clockwise in the lower cortex and clockwise in the upper cortex (Fig.~\ref{fig:stochastic_ex}(c)). At the stall force (when $\langle{y}^\pm_{\textrm{b}}\rangle\approx1$), the direction of the solution loop is determined by the direction of acceleration of the spindle pole with respect to the cortex. That is, a force generator in the lower cortex whose elastic linker is at ${y}^{(n)\pm}_{\textrm{b}}=1$ will be decreasing its extension as the spindle pole accelerates toward it (negative acceleration, green point in Fig.~\ref{fig:stochastic_ex}(a,c)) and increasing as the spindle pole accelerates away (positive acceleration, yellow point  in Fig.~\ref{fig:stochastic_ex}(a,c)). 

Removing the tension sensitivity of unbinding by setting $\gamma=0$ results in less well-defined oscillations (Fig.~\ref{fig:stochastic_ex}(d)) of reduced amplitude relative the baseline case shown in Fig.~\ref{fig:stochastic_ex}(a). Thus the tension-sensitive unbinding rate appears to promote coherent oscillations of the spindle pole, although fluctuations persist due to the stochastic binding and unbinding. 
Similarly, increasing the restoring force by increasing the parameter $K$ reduces the deviation in the position of the spindle pole from the centre (Fig.~\ref{fig:stochastic_ex}(e)), but also leads to a marked reduction in the coherence of the spindle motion (compare Fig.~\ref{fig:stochastic_ex}(a) and (e)). 

\section{A Fokker--Planck description}
\label{sec:FP}

Simulating the system stochastically reveals the role of noise in individual realisations of spindle dynamics.  To explore properties of the model over multiple realisations in a more computationally efficient manner, we turn to a system of partial differential equations (PDEs) for probability density functions (pdfs) $P_{\mathrm{b}(\mathrm{u})}({y},{t})$ for the extensions of bound and unbound linkers at either cortex, where the elastic linker extension ${y}$ is now considered as a continuous variable.  The model may be written as
\begin{subequations} 
\begin{align}\label{eq:FPEb}
    {P}_{\textrm{b},{t}}^\pm+{J}_{\textrm{b},{y}}^\pm&={\omega}_{\textrm{on}}{P}_{\textrm{u}}^\pm-{\omega}_0e^{\gamma{y}}{P}_{\textrm{b}}^\pm, &{J}_{\textrm{b}}^\pm &= {v}^\pm_{\textrm{b}}{P}_{\textrm{b}}^\pm - {D_{\mathrm{b}}}{P}_{\textrm{b},{y}}^\pm,  \\
    {P}_{\textrm{u},{t}}^\pm+{J}_{\textrm{u},{y}}^\pm&=-{\omega}_{\textrm{on}}{P}_{\textrm{u}}^\pm+{\omega}_0e^{\gamma{y}}{P}_{\textrm{b}}^\pm,  &  {J}_{\textrm{u}}^\pm &= v_u^\pm {P}_{\textrm{u}}^\pm - \Gamma{D_{\mathrm{u}}}{P}_{\textrm{u},{y}}^\pm,
    \label{eq:FPEu}
\end{align}
\label{eq:FPE}
\end{subequations}
where
\begin{equation}\label{eq:FPEvel}
    {v}^\pm_{\textrm{b}}=1-{y}\mp\frac{\textrm{d}{z}}{\textrm{d}{t}}, \quad {v}^\pm_{\textrm{u}}=-\Gamma{y}.
\end{equation}
Equation (\ref{eq:FPE}) is a nondimensional version of the {\color{black}mean-field} Fokker--Planck equations proposed by \cite{grill2005theory}.  The continuous velocities ${v}^\pm_{\textrm{b(u)}}({y})$ in (\ref{eq:FPEvel}) evolve as in (\ref{eq:velocity2_nondim}).   The pulling force toward each cortex (\ref{eq:popforcing}) is calculated as
\begin{equation}\label{eq:FPEForce}
    {F}^\pm = N\int^{{y}_{\textrm{max}}}_{0} {y}{P}_{\textrm{b}}^\pm({y},{t})\textrm{d}{y},
\end{equation}
modifying the force balance on the spindle (\ref{eq:polepos_nondim}), which becomes 
\begin{equation} 
    {\xi}{z}_{{t}} = -K{z}-N\left(\int_0^{{y}_{\textrm{max}}} {y} {P}^{-}_{\textrm{b}}\textrm{d}{y}-\int_0^{{y}_{\textrm{max}}} {y} {P}^{+}_{\textrm{b}}\textrm{d}{y}\right). \label{eq:poleeom}
\end{equation}
The boundary conditions
\begin{equation}\label{eq:Jbounds}
    {J}^\pm_{\textrm{b}}
    = {J}^\pm_{\textrm{u}}
    =0\quad \mathrm{at}~ {y}=0~ \mathrm{and}~ {y}={y}_{\max},
\end{equation}
ensure conservation of total probability
\begin{align}
    \int_0^{{y}_{\textrm{max}}} \left({P}_{\textrm{b}}^\pm+{P}_{\textrm{u}}^\pm\right)\textrm{d}{y} = 1. \label{eq:Pu+Pb=1}
\end{align}
Given some initial conditions ${P}_{\textrm{b}}^\pm\left({y},0\right)={P}_{\textrm{b}0}^\pm\left({y}\right)$, ${P}_{\textrm{u}}^\pm\left({y},0\right)={P}_{\textrm{u}0}^\pm\left({y}\right)$, and ${z}\left(0\right)={z}_{0}$, the system (\ref{eq:FPE}--\ref{eq:Jbounds}) 
may be solved in time to return the dynamics of the spindle pole and the populations of cortical force generators, represented as probability densities over multiple realisations of the system. We computed numerical solutions using the method of lines.


\begin{figure}[h!]
    \centering
    \includegraphics[width=0.9\textwidth]{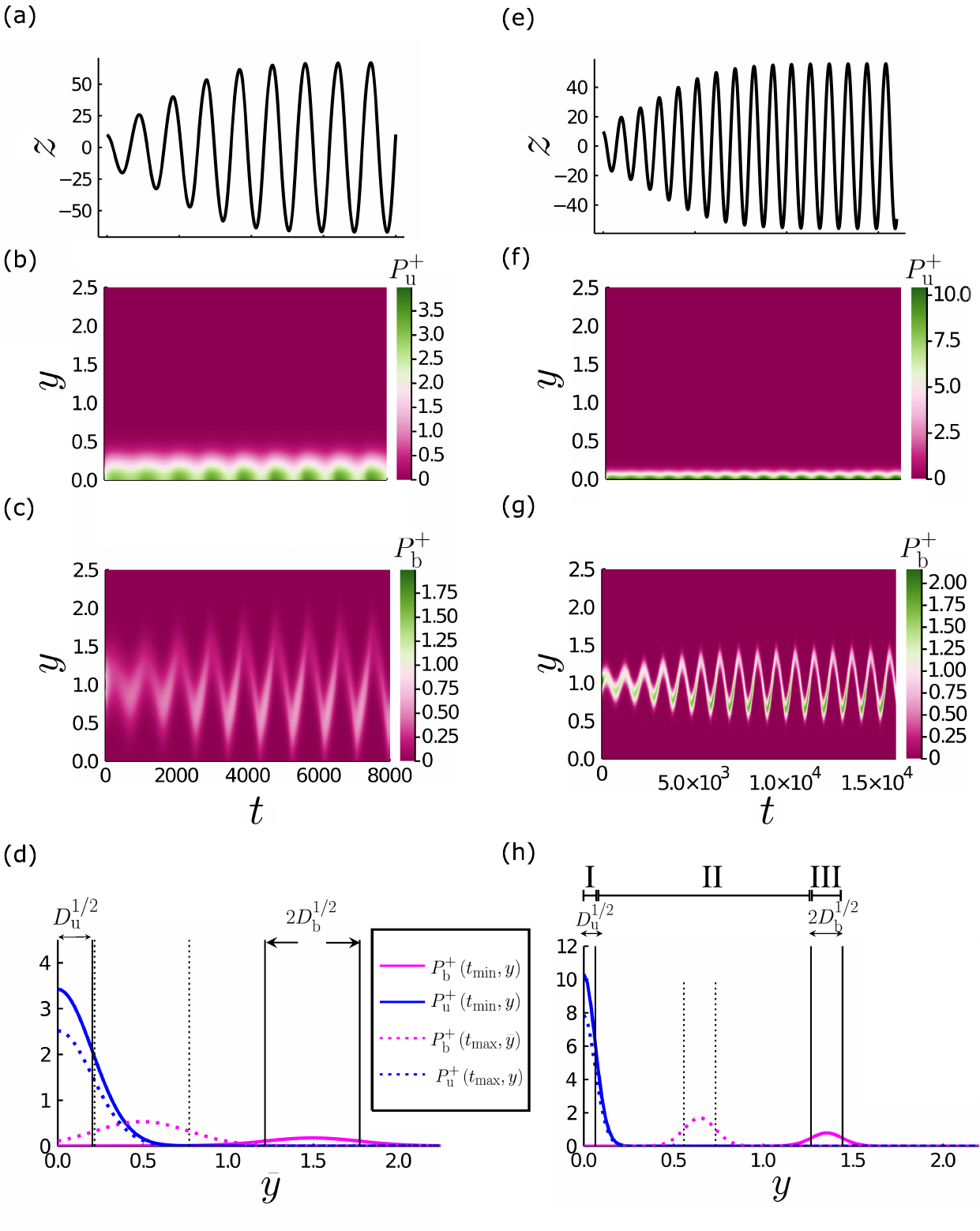}
    \caption{{\textbf{The effect of varying the magnitude of diffusion in the Fokker--Planck description}}. (a,e) Example solution to equations (\ref{eq:FPEb}, \ref{eq:FPEu}, \ref{eq:poleeom}), showing the pole position, ${z}$ versus time ${t}$. Diffusion parameters ${D_{\mathrm{b}}}, {D_{\mathrm{u}}}$ are a factor of 10 smaller in the right column than in the left column. (b,f) Heat map of ${P}_{\textrm{u}}^+(y,t)$. (c,g) Heat map of ${P}_{\textrm{b}}^+(y,t)$. (d,h) Probability density functions in the upper cortex at two instances of time. Solid lines: ${t} = {t}_{\textrm{min}}$, when the spindle pole is at ${z}=0$ and moving toward its minimum value (${z}_{{t}}<0$). Dotted lines: ${t} = {t}_{\textrm{max}}$, when the spindle pole is at ${z}=0$ and moving toward its maximum value (${z}_{{t}}>0$). The peak widths scale with ${D_{\mathrm{u}}}^{1/2}$ and ${D_{\mathrm{b}}}^{1/2}$ as indicated. (h) The three regions used to reduce the system of PDEs to ODEs are indicated by roman numerals I, II, and III. The behaviour of the pdfs in the lower cortex are in antiphase to the behaviour seen here. Solutions were obtained using parameters as in Table~\ref{table:nondims} plus: $N=25$; (a-d) baseline diffusivities ${D_{\mathrm{b}}} = 0.08$, ${D_{\mathrm{u}}}=0.04$; (e-h) ${D_{\mathrm{b}}} = 8\times 10^{-3}$, ${D_{\mathrm{u}}}=4\times 10^{-3}$.}
    \label{fig:PDEsol}
\end{figure}

{\color{black}The underlying stochastic system (Appendix~\ref{secA1}) combines two sets of random processes: binding and unbinding of force generators; and the random walk of force generators along microtubules.  The description of the system provided by (\ref{eq:FPE}) can be described as mean-field in the sense that it combines a deterministic model of binding/unbinding kinetics (via the reaction terms in (\ref{eq:FPE})) with a stochastic description of force-generator motion (via the diffusive terms in (\ref{eq:FPE})).  We shall revisit the role of randomness in binding kinetics in Section~\ref{sec:noisosc} below.}

The solutions {\color{black}of (\ref{eq:FPE})--(\ref{eq:Pu+Pb=1})} presented in Fig.~\ref{fig:PDEsol} show an oscillating spindle displacement ${z}(t)$ corresponding to fluctuations in ${P}^{\pm}_{\textrm{b}}\left(y,{t}\right)$ and ${P}^{\pm}_{\textrm{u}}\left(y,{t}\right)$.
For large $\Gamma$, (\ref{eq:FPEu}) is dominated by the advective term which sweeps any unbound force generators with a non-zero extension down toward ${y}=0$. As there is no flux through this boundary by (\ref{eq:Jbounds}), ${P}_{\textrm{u}}^\pm$ has a defined peak at ${y}=0$ which decays with ${y}$ over the diffusive lengthscale $D_{\mathrm{u}}^{1/2}$.  For the bound pdfs ${P}_{\textrm{b}}^\pm$, the location ${y}_{\textrm{c}}^\pm$ and amplitude ${P}_{\textrm{b}}^{\pm,\textrm{max}}$ of the maximum of the pdf oscillate concurrently with ${z}$ (Fig.~\ref{fig:PDEsol}(c, g)), mirroring the behaviour of the average extension $\langle{y}^\pm_{\textrm{b}}\rangle$ and number of bound force generators $n^\pm_{\textrm{b}}$ in the stochastic simulation (Fig.~\ref{fig:stochastic_ex}(b)).  Variations of the initial conditions ${P}^{\pm}_{\textrm{b0}}$ and ${P}^{\pm}_{\textrm{u0}}$ had no effect on the final solutions following initial transients (data not shown). 
 
Decreasing ${D_{\mathrm{b}}}$ and ${D_{\mathrm{u}}}$ by a factor of $10$ results in taller and narrower pdfs (Fig.~\ref{fig:PDEsol}(d,h)),  
confining ${P}_{\textrm{b}}^\pm$ to a region of ${y}$ which is spatially separated from ${P}_{\textrm{u}}^\pm$ at all times. In this limit we can partition the ${y}$-domain into three distinct regions: I, of width $\mathcal{O}(D_{\mathrm{u}}^{1/2})$, encompassing the peak of ${P}_{\textrm{u}}^+$; III, of width $\mathcal{O}(D_{\mathrm{b}}^{1/2})$, encompassing the peak of ${P}_{\textrm{b}}^+$; and II between them, which remains distinct throughout an entire oscillation (Fig.~\ref{fig:PDEsol}(h), see Online Resource 2c).  As well as modulating the shape of the pdfs, ${D_{\mathrm{b}}}$ and ${D_{\mathrm{u}}}$ also affect the resulting dynamics of the spindle pole. Decreasing ${D_{\mathrm{b}}}$ and ${D_{\mathrm{u}}}$ results in an increased period, $T$, of oscillation ($T\approx 890$ increases to $T\approx 1000$ upon a decrease in ${D_{\mathrm{b}}}$ and ${D_{\mathrm{u}}}$ by a factor of 10), a decrease in the amplitude of the oscillation (Fig.~\ref{fig:PDEsol}(a, e)) and longer transients 
(Fig.~\ref{fig:PDEsol}(a, e)).   


 \begin{figure}
    \centering
    \includegraphics[width=\textwidth]{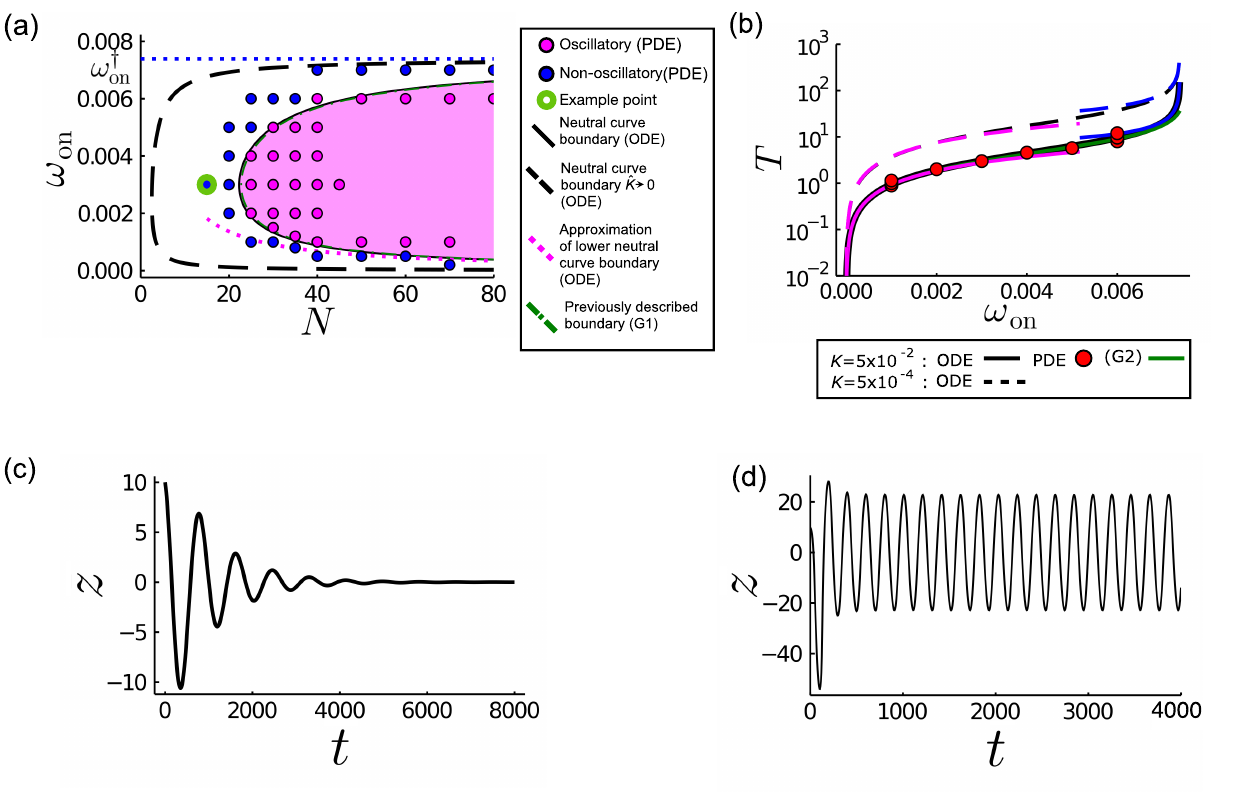}
    \caption{\textbf{The stability boundary between oscillatory and non-oscillatory solutions is affected by the magnitude of diffusive terms.} (a) Numerically solving the Fokker--Planck system (circles) reveals a boundary in ($N,{\omega}_{\textrm{on}}$) space which separates oscillatory from non-oscillatory solutions. Each circle represents a numerical solution, labelled magenta if the spindle pole has sustained oscillations and blue if the spindle pole position decays to $z=0$ for large $t$. The point with the green boundary is the location in parameter space at which the solutions (c) and (d) sit. Other parameters are as in Table~\ref{table:nondims} except that ${D_{\mathrm{b}}} = 8\times 10^{-3}$ and ${D_{\mathrm{u}}}=4\times 10^{-3}$. {\color{black}For later reference,} the shaded magenta area represents the region where oscillatory solutions exist as determined by stability analysis of the ODEs (\ref{eq:stab}) using equivalent parameters. The dashed curve (black) shows the same threshold in the limit of {\color{black}weak restoring force} ($\hat{K}\rightarrow0$, {\color{black} see (\ref{eq:odeparams})}) determined by (\ref{eq:smallKstep}). The dotted magenta curve shows the asymptote of the lower boundary for $N\gg1$ and $N\omega_{\mathrm{on}}=\mathcal{O}\left(1\right)$, as in (\ref{eq:lower}).  The dashed green curve shows the stability boundary (G1) predicted by  (\ref{eq:grillK}) from \cite{grill2005theory}.  (b) The relationship between the period of oscillation and the binding rate ${\omega}_{\textrm{on}}$ using (\ref{eq:ODEfreq}), along the neutral stability curve (\ref{eq:stab}). The period is unbounded as ${\omega}_{\textrm{on}}\rightarrow{\omega}_{\textrm{on}}^\dag$.  Points denote the periods taken from PDE solutions along the approximate neutral curve identified in Fig.~\ref{fig:PDE_N15}(a). The magenta curves represent the approximations to the period for small 
    $\bar{\omega}_{\textrm{on}}$ as in (\ref{eq:tlower}). The blue curves represent the approximations to the period as $\omega_{\textrm{on}}\rightarrow\omega_{\textrm{on}}^{\dag}$ as in (\ref{eq:tupper}). The dashed green curve shows the period (G2) predicted by  (\ref{eq:grillomega}) from \cite{grill2005theory}. (c) Spindle pole position ${z}$ in time ${t}$ at the example point (green in (a)), from a PDE solution. (d) A PDE solution replicating (c) except that ${D_{\mathrm{b}}} = 8\times10^{-1}$ and ${D_{\mathrm{u}}} = 4\times10^{-1}$.}
    \label{fig:PDE_N15} 
\end{figure}

The solution in Fig.~\ref{fig:PDEsol}(a-d) was run with parameters matching those in the stochastic simulation in Fig.~\ref{fig:stochastic_ex}, except that $N=25$ in the former and  $N=15$ in the latter.  Nevertheless, PDE predictions show a comparable period, without capturing the detailed fluctuations in an individual realisation.  To obtain a broader view of parameter dependence, solutions of (\ref{eq:FPE}--\ref{eq:Jbounds}) for a range of values of $N$ and ${\omega}_{\textrm{on}}$ are reported in Fig.~\ref{fig:PDE_N15}(a).  For the baseline value $\omega_{\mathrm{on}}=0.003$, sustained oscillations arise in the PDE model with $N=25$ and low diffusivities (as in Fig.~\ref{fig:PDEsol}(e-h)) but not $N=15$ (see Fig.~\ref{fig:PDE_N15}(c)).  Increasing diffusivities with $N=15$ leads to the sustained oscillations seen in the PDE model in Fig.~\ref{fig:PDE_N15}(d). Noise, therefore, is likely to play a role in promoting oscillatory dynamics.

The PDE stability boundary for low diffusivities is mapped out in ($N$, ${\omega}_{\textrm{on}}$)-space (Fig.~\ref{fig:PDE_N15}(a)), distinguishing oscillatory from non-oscillatory solutions. The period of oscillation for neutrally-stable disturbances increases with $\omega_{on}$ (Fig.~\ref{fig:PDE_N15}(b)).  Reduction of the number of force generators, leading to a decrease in pulling forces, results in a cessation of oscillations (Fig.~\ref{fig:PDE_N15}(a,c)). For large $N$, two thresholds exist for values of ${\omega}_{\textrm{on}}$ at which oscillations arise, with the oscillatory section of parameter space forming a wedge shape
(Fig.~\ref{fig:PDE_N15}(a)). 
We explore the origins of these thresholds in more detail below. This wedge-shaped parameter space was described previously by \cite{grill2005theory} through analysis of a reduced model where it was assumed that unbound force generators instantaneously relax down to zero extension. The presence of a threshold between oscillatory and non-oscillatory solutions has been experimentally validated in \emph{C. elegans} embryos \citep{pecreaux2006spindle}.


Significantly, the oscillatory behaviour reported in stochastic simulations for $N=15$ (Fig.~\ref{fig:stochastic_ex}(a)), lies in a regime in which the Fokker--Planck model predicts steady distributions of $P_{b(u)}^{\pm}$ with $z\rightarrow 0$ at large times (Fig.~\ref{fig:PDE_N15}(c)).  We provide evidence below that the sustained oscillations in Fig.~\ref{fig:stochastic_ex}(a) are noise-driven (or a form of stochastic resonance \citep{erban2020}), {\color{black} with noise arising from stochastic binding kinetics, a feature missing from the mean-field Fokker--Planck model.}

\begin{figure}
    \centering
    \includegraphics[width=0.85\textwidth]{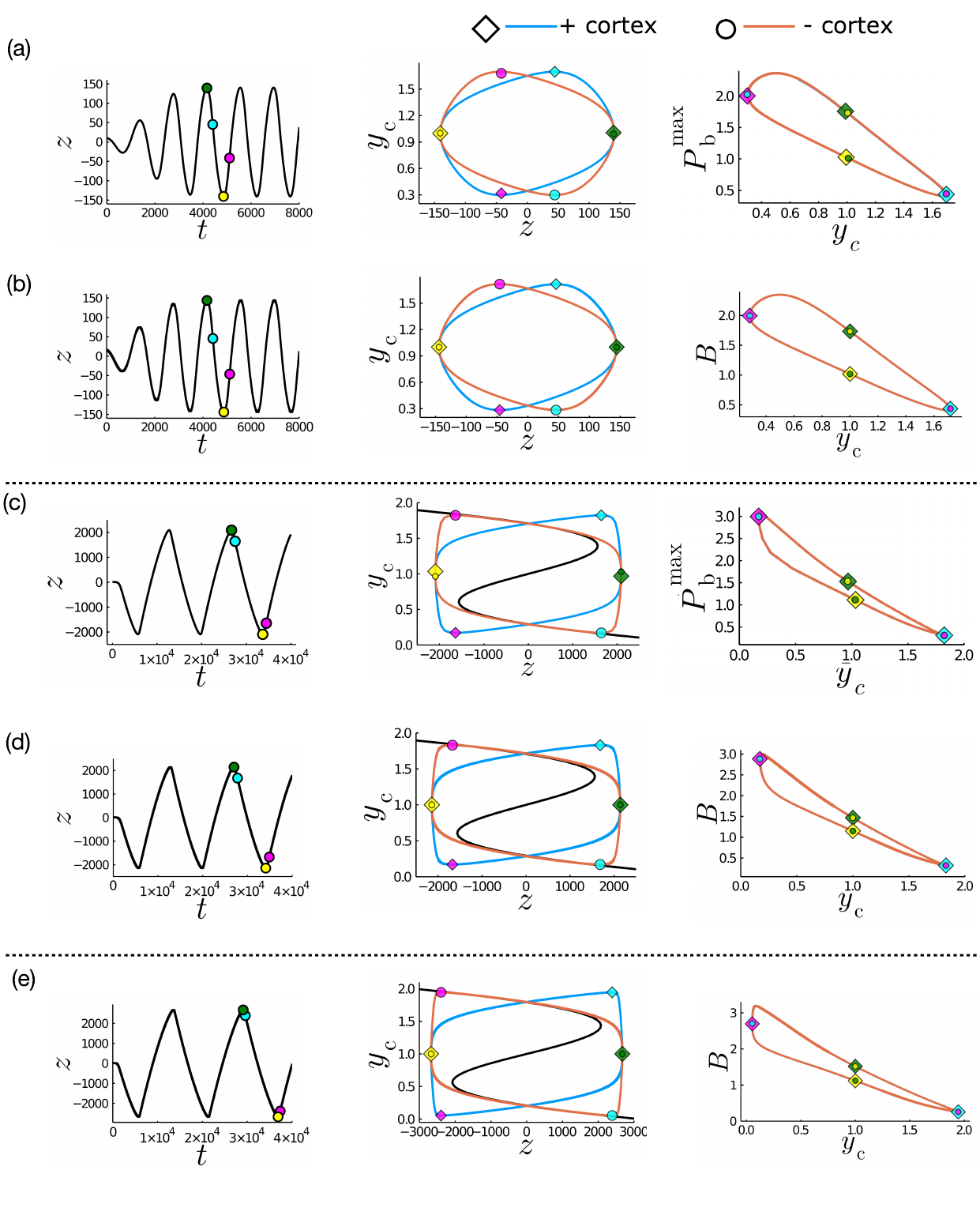}
    \caption{{\textbf{Comparison of PDE (a, c) and ODE ( solutions for equivalent parameters.}} PDE and ODE solutions for equivalent parameters are presented, with non-equivalent solutions separated by a dotted line. (a, c) solutions of the PDEs; (b, d, e) solutions of the ODEs. First column: spindle pole position ${z}$. Second column: centre of the bound pdf as a function of pole position ${y}^\pm_{\textrm{c}}\left({z}\right)$. Third column: amplitude of the bound pdf as a function of the location of its peak (${P}^\pm_{\textrm{b}}\left({y}_{\textrm{c}}\right)$ for PDE solutions (a,c); $B^\pm=\hat{B}^\pm/\sqrt{2\pi{D_{\mathrm{b}}}}$ for ODE solutions (b, d, e). PDE solutions were obtained using parameters are as in Table~\ref{table:nondims} except ${D_{\mathrm{b}}}=8\times10^{-3}$, ${D_{\mathrm{u}}}=4\times10^{-3}$, (a)  $N=45$ and (c) $K=5\times10^{-4}$ and $N=15$.  ODE solutions obtained using (b) equivalent parameters to (a); (d) equivalent parameters to (c); and (e) Equivalent parameters to (a) with $N=1500$. Line colours correspond to solutions in each cortex (blue = upper, orange = lower). The black curves in the centre column represent the predicted limit cycle as $\hat{K}\rightarrow0$, as determined by the inversion of (\ref{eq:branch}). Scatterpoints denote the positions of: maximum amplitude (blue $z>0$, yellow $z<0$), maximum spindle pole velocity (cyan $z_{\textrm{t}}<0$, magenta $z_{\textrm{t}}>0$.}
    \label{fig:ODESol}
\end{figure}

Additional PDE solutions with low diffusivities are reported in Fig.~\ref{fig:ODESol}(a,c), illustrating respectively the impact of increasing $N$ and decreasing the strength of the restoring force $K$.  The latter leads to larger-amplitude oscillations with a relaxation structure, 
characterised by periods of approximately uniform spindle velocity, interspersed with rapid changes in direction (Figs \ref{fig:ODESol}(c)). Correspondingly, the oscillations combine slow phases in the time-evolution of ${P}^\pm_{\textrm{b}}$ and ${y}^\pm_{\textrm{c}}$ (Fig.~\ref{fig:ODESol}(c)) in which $z$ is approximately linear in $t$, with short intervals in which the rapid change in the direction of motion of the spindle pole coincides with fast changes in the extension of the force generator bound probability centre ${y}^\pm_{\textrm{c}}$ and amplitude ${P}^\pm_{\textrm{b}}$.  We explore the origins of this strongly nonlinear behaviour below, through comparison to a simplified model reported in the remaining panels of Fig.~\ref{fig:ODESol}.

The Fokker--Planck description (\ref{eq:FPE}--\ref{eq:Pu+Pb=1}) reveals many of the characteristics promoting spindle-pole oscillations, but still requires extensive computation.  We now reduce this model to a system of ODEs by asymptotic analysis, allowing us to more fully explore the relationships between the most important factors promoting oscillations.  Rather than follow the heuristic approach proposed by \cite{grill2005theory}, we seek a systematic reduction valid at an appropriate distinguished limit in parameter space.

\section{Asymptotic reduction to ODEs}
\label{sec:odes}

When diffusivities are small, the PDEs reveal distinct regions of ${y}$ space where the pdfs ${P}^\pm_{\textrm{u}}$ and ${P}^\pm_{\textrm{b}}$ have most of their mass (Fig.~\ref{fig:PDEsol}(h)). While varying the diffusive terms has an impact on the oscillations, with larger diffusive terms promoting oscillations (Fig.~\ref{fig:PDE_N15}(d)), the amplitudes and periods of the more and less diffusive solutions are still of a similar order. We now pursue the behaviour of the model with lower diffusivity to create a system of ODEs. 

We develop an approximation to the oscillating spindle system in a distinguished limit for which {\hbox{${\omega}_{\textrm{on}}\sim{\omega}_0\sim{D_{\mathrm{b}}}^{1/2}\sim{D_{\mathrm{u}}}^{1/2}\ll1$}} (where $\sim$ means `scales like') and rescale the time and spindle position parameters in the PDE problem (\ref{eq:FPE}), (\ref{eq:FPEvel}) and (\ref{eq:poleeom}) by ${t}=\tilde{t}/{\omega}_{\textrm{on}}$ and ${z}=\tilde{z}/{\omega}_{\textrm{on}}$.  {\color{black}In other words, we assume that binding kinetics happens slowly relative to movement of force generators, and that fluctuations of such movement are weak.} The range of extension values ${y}$ are split into the three regions identified in Fig.~\ref{fig:PDEsol}(h): I, over which ${P}_{\textrm{u}}^\pm$ is peaked around ${y}=0$ with a width ${D_{\mathrm{u}}}^{1/2}$; III, over which ${P}_{\textrm{b}}^\pm$ is peaked with a width of ${D_{\mathrm{b}}}^{1/2}$ but whose centre moves as ${y}_{\textrm{c}}=1\mp\tilde{z}_{\tilde{t}}$; and II, where advective terms dominate.

In Appendix~\ref{A2}, we express the governing equations in rescaled coordinates, expand in powers of $\omega_{\mathrm{on}}$, solve for
${P}_{\textrm{u}}^\pm$ in region I and ${P}_{\textrm{b}}^\pm$ in region III, and then match asymptotic limits across region II.  This procedure yields the three coupled ODEs
\begin{subequations}
\label{eq:ODE}
\begin{align}
    \left(\hat{\xi}+\hat{B}^++\hat{B}^-\right)\tilde{z}_{\tilde{t}}+\hat{K}\tilde{z}&=\hat{B}^+-\hat{B}^-, \label{eq:ODE1} \\
    \left(1+\rho e^{\gamma\left(1\mp\tilde{z}_{\tilde{t}}\right)}\right)\hat{B}^\pm+\hat{B}^\pm_{\tilde{t}} &= 1. \label{eq:ODE2}
    \end{align}
\end{subequations}
Here $\hat{B}^\pm\left(\tilde{t}\right)=\sqrt{2\pi{D_{\mathrm{b}}}}B^\pm$, where $B^\pm$ approximates the amplitude of the peak of ${P}^\pm_{\textrm{b}}$. 
The parameters in (\ref{eq:ODE}) are 
\begin{equation}
\label{eq:odeparams}
\rho=\frac{{\omega}_0}{{\omega}_{\textrm{on}}}=\frac{\bar{\omega}_0}{\bar{\omega}_{\textrm{on}}}, \quad
    \hat{\xi}=\frac{\xi}{N}=\frac{\bar{\xi}v_0}{f_0 N}, \quad \hat{K}=\frac{K}{N{\omega}_{\textrm{on}}}=\frac{k_{\mathrm{MT}}v_0}{N \bar{\omega}_{\mathrm{on}} f_0}.
\end{equation}
Formally, the parameters (\ref{eq:odeparams}) are assumed to remain $\mathcal{O}(1)$ in the limit ${\omega}_{\textrm{on}}\sim{\omega}_0\sim{D_{\mathrm{b}}}^{1/2}\sim{D_{\mathrm{u}}}^{1/2}\ll 1$.  $\rho$ is the binding affinity (under zero load) of force generators for microtubules; $\hat{\xi}$ measures the spindle drag (assuming the spindle moves at the walking speed of a linker) relative to the stall force generated by the full population of linkers; $\hat{K}$ measures the restoring force, driving the spindle to the centre of the cell (assuming a displacement of the spindle comparable to the distance walked by a linker), relative to the stall forces generated by the full population of linkers.  We solved (\ref{eq:ODE}) numerically, imposing initial conditions $\tilde{z}_0$ and $\hat{B}^\pm_0$.  

The ODEs defined by \cite{grill2005theory} may be re-written in the notation used above as
\begin{subequations}
\label{eq:GrillODEs}
    \begin{align}
        \hat{\xi}\tilde{z}_{\tilde{t}} + \hat{K}  \tilde{z} &= \hat{B}^+\tilde{y}^+-\hat{B}^-\tilde{y}^- \label{eq:GrillODE1} \\
        \left(1+\rho e^{\gamma\tilde{y}^\pm}\right)\hat{B}^\pm + \hat{B}^\pm_{\tilde{t}} &= 1 \label{eq:GrillODE2}
    \end{align}
where $\tilde{y}^\pm$, the typical length a linker extends before it detaches, is determined from
\begin{equation}\label{eq:GrillY}
    \mp\tilde{z}_{\tilde{t}}=\left(\omega_0e^{\gamma\tilde{y}^\pm}+1\right)\tilde{y}^\pm-1.
\end{equation}
\end{subequations}
Assuming $\omega_0\ll1$ in (\ref{eq:GrillY}), then $\tilde{y}\simeq1\mp\tilde{z}_{\tilde{t}}$ and we recover (\ref{eq:ODE}).  Our asymptotic reduction therefore recovers a special case of the heuristically-determined ODEs presented by \cite{grill2005theory}.


In Fig.~\ref{fig:ODESol}(a-d), two solutions of the Fokker--Planck system are compared with solutions of the ODEs (\ref{eq:ODE}) for equivalent parameters, taking ${y}_{\textrm{c}}^\pm=1\mp\tilde{z}_{\tilde{t}}$. The spindle pole dynamics and associated force generator behaviours in both cortices are fully captured in the ODE model in both cases.

Relaxation oscillations arise when pulling forces dominate over restoring forces, through a reduction of $\hat{K}$.
Fig.~\ref{fig:ODESol}(c,d) shows a strong match between the relaxation oscillations returned by the PDE and ODE models for an equivalent reduction in $K$, with similar period and amplitude as well as shape. Likewise applying an increase in $N$ (Fig.~\ref{fig:ODESol}(e)), the ODE model predicts a strong relaxation structure 
(with very sharp changes in ${y}_{\textrm{c}}^\pm$ at the peaks of oscillation). {\color{black}{The amplitude of this relaxation structure is, however, larger than would be viable within a biological cell. This model description omits explicit boundaries, and we would expect that within a cell the presence of a boundary would introduce a nonlinear contribution to the restoring force close to the cell edge. Despite this, we}}
conclude that the balance of pulling to restoring forces controls the {\color{black}{general}} structure of the oscillation of the spindle pole.

Interestingly, the oscillations in the lower-$\hat{K}$ and high-$N$ relaxation oscillation are slightly different. For low $\hat{K}$, the peak of the bound pdf hits its maximum at the same time as the spindle pole experiences its maximum velocity (when ${y}_{\textrm{c}}^\pm$ is at its minimum value, Fig.~\ref{fig:ODESol}(c,d)). Alternatively, when $N$ is increased (leading to changes in $\hat{K}$ and $\hat{\xi}$), the maximum of the peak of the bound pdf lags behind the spindle pole velocity (Fig.~\ref{fig:ODESol}(e)). This lag 
represents a delay between the binding of the force generators and the movements of the spindle pole, likely due to there being a greater number of force generators in the system to bind to the microtubules before saturation of the force generators onto the microtubule. Indeed when ${\omega}_{\textrm{on}}$ is small, the oscillations of the spindle pole are more non-linear (data not shown), as the number of bound force generators takes longer to saturate.

\subsection{Stability analysis}

The simplicity of the ODE model (\ref{eq:ODE}) lends itself to stability analysis.  Linearising about the steady state
\begin{equation}\label{eq:Bsteady}
    \hat{B}^{*\pm} = \lambda^{-1}, \quad 
        \tilde{z}^* = 0, \quad \lambda\equiv 1+\rho e^\gamma,
\end{equation}
assuming that small disturbances are proportional to $e^{s\tilde{t}}$, yields the relationship
\begin{equation}
    \left(s+\lambda\right)\left(s^2+s\left(\lambda+\frac{\hat{K}\lambda}{\lambda\hat{\xi}+2}-\frac{2\gamma\left(\lambda-1\right)}{\lambda\hat{\xi}+2}\right)+\frac{\hat{K}\lambda^2}{\lambda\hat{\xi}+2}\right)=0.
    \label{eq:cubic}
\end{equation}
Setting $s=\mu+i\Omega$ and collecting real and imaginary parts defines the growth rate
\begin{subequations}
\begin{equation}\label{eq:growthrate}
    \mu = \frac{2\gamma\left(\lambda-1\right)-\hat{K}\lambda-\lambda\left(\lambda\hat{\xi}+2\right)}{2\left(\lambda\hat{\xi}+2\right)}
\end{equation}
and frequency of oscillation 
\begin{equation}\label{eq:fullperiod}
    \Omega^2 = \frac{\hat{K}\lambda^2}{\lambda\hat{\xi}+2}-\frac14\left(\frac{\hat{\xi}\lambda+\hat{K}-2\left(\frac\gamma\lambda\left(\lambda-1\right)-1\right)}{\hat{\xi}+2/\lambda}\right)^2.
\end{equation}
\end{subequations}
Setting $\mu=0$ at the onset of neutral oscillations identifies the frequency
\begin{subequations}
\begin{equation}\label{eq:ODEfreq}
    \Omega^2=\frac{\hat{K}\lambda^2}{2+\hat{\xi}\lambda} 
\end{equation}
and the stability threshold
\begin{equation}
    \hat{K}=2\left(\frac{\gamma}{\lambda}\left(\lambda-1\right)-1\right)-\hat{\xi}\lambda.
    \label{eq:stab}
\end{equation}
\end{subequations}
Both (\ref{eq:stab}) and (\ref{eq:ODEfreq}) provide good predictions of the stability boundary identified by PDE solutions (Fig.~\ref{fig:PDE_N15}(a)) and the period of oscillations at the stability boundary (Fig.~\ref{fig:PDE_N15}(b)).

The frequency of oscillation determined by \cite{grill2005theory} via (\ref{eq:GrillODEs}) may be rewritten using the notation above as
\begin{subequations}
\begin{equation}\label{eq:grillomega}
    \Omega^2 = \frac{\hat{K}\lambda^2\left(1+\left(\gamma+1\right)\omega_0 e^{\gamma}\right)}{\hat{\xi}\lambda\left(1+\left(\gamma+1\right)\omega_0 e^\gamma\right)+2}
\end{equation}
at the stability threshold
\begin{equation}\label{eq:grillK}
    \hat{K}=\frac{2\left(\frac{\gamma}{\lambda}\left(\lambda-1\right)-1\right)}{\omega_{\textrm{on}}\left(\lambda-1\right)\left(\gamma+1\right)+1}-\hat{\xi}\lambda.
    \end{equation}
\end{subequations}
Equations (\ref{eq:grillomega}) and (\ref{eq:grillK}) are equivalent to (\ref{eq:ODEfreq}) and ({\ref{eq:stab}}) respectively when $\omega_0\sim\omega_{\textrm{on}}\ll1$. (\ref{eq:stab}) and (\ref{eq:grillK}) are compared in Fig.~\ref{fig:PDE_N15}(a), showing a near-perfect match despite the additional terms present in (\ref{eq:grillK}); both predictions bound almost perfectly the oscillatory region observed by individual solutions of the PDEs. 

We highlight two limits of (\ref{eq:stab}).  First, taking $\hat{K}\rightarrow 0$, leaving $\hat{\xi}={\xi}/N$ as the only parameter which depends on $N$, (\ref{eq:stab}) reduces to 
\begin{equation}\label{eq:smallKstep}
    N=\frac{{\xi}\lambda^2}{2\gamma\left(\lambda-1\right)-2\lambda}=\frac{\xi(\omega_{\mathrm{on}}+\omega_0 e^\gamma)^2}{2\omega_{\mathrm{on}}[\omega_0 e^\gamma(\gamma-1)-\omega_{\mathrm{on}}]},
\end{equation}
provided the denominator $2[\gamma(\lambda-1)-\lambda]$ is positive, \hbox{i.e.}
\begin{equation}
    {\omega}_{\textrm{on}}<\omega_{\mathrm{on}}^\dag \equiv \left(\gamma-1\right){\omega}_0 e^{\gamma}.
    \label{eq:omedag}
\end{equation}
In this limit, the period of oscillation $T=2\pi/\Omega$ is 
\begin{equation}
    T\approx 2\pi\left(\frac{\left(\gamma-1\right)N \omega_{\mathrm{on}}^\dag}{K\gamma} \right)^{1/2}.
    \label{eq:tupper}
\end{equation}
$\omega_{\mathrm{on}}\rightarrow \omega_{\mathrm{on}}^\dag$ also appears in the large-$N$ limit of (\ref{eq:stab}), taking  $\hat{K}\sim\hat{\xi}\ll 1$.  In addition, taking $N\gg 1$, $\omega_{\mathrm{on}}\ll 1$ with $N \omega_{\mathrm{on}}=O(1)$, we recover the additional approximation (provided $\gamma>1$)
\begin{equation}
    \omega_{\mathrm{on}}\approx \frac{K+\xi \omega_0 e^\gamma}{2N(\gamma-1)}
    \label{eq:lower}
\end{equation}
for which
\begin{equation}
    T=2\pi \frac{\omega_{\mathrm{on}}}{\omega_{\mathrm{on}}^\dag} \left[\frac{(\gamma-1)(K+2\xi \omega_0 e^\gamma)}{K} \right]^{1/2}.
    \label{eq:tlower}
\end{equation}
Thus, for large $N$, the upper branch of the stability boundary defined by (\ref{eq:stab}) in Fig.~\ref{fig:PDE_N15}(a) approaches $\omega_{\mathrm{on}}=\omega_{\mathrm{on}}^\dag$ in (\ref{eq:omedag}), confirming that a necessary condition for oscillations is that the tension-sensitivity parameter satisfies $\gamma>1$, \hbox{i.e.} that linkers exhibit slip-bond behaviour.
Indeed, removal of the tension-sensitivity of the unbinding rate in the stochastic simulations leads to a reduction of the coherence of the oscillatory behaviour of the spindle pole (Fig.~\ref{fig:stochastic_ex}(d)).  The upper-branch asymptote $\omega_{\mathrm{on}}=\omega_{\mathrm{on}}^\dag$ appears to be shared also by PDE solutions (which suggests an upper stability threshold between $0.006<{\omega}_{\textrm{on}}^\dag<0.007$ for $N\leq80$, within $80\%$ of ${\omega}_{\textrm{on}}^\dag = 0.0074$).   Also in the large-$N$ limit, the lower branch of (\ref{eq:stab}) is captured by (\ref{eq:lower}), consistent with PDE solutions in this limit.  This limit shows explicitly how increasing the restoring force $K$ has a stabilising effect.

We also recall that, in the Fokker--Planck model, decreasing the restoring force parameter $K$ promotes oscillations at smaller $N$ (Fig.~\ref{fig:ODESol}(c), where $N=15$).  This behaviour is conserved in the ODE system, where the low-$\hat{K}$ approximation (\ref{eq:smallKstep})
shown in Fig.~\ref{fig:PDE_N15}(a), predicts oscillations in a greater region of the ($N, {\omega}_{\textrm{on}}$)-plane.  Evaluating $\mathrm{d}\omega_{\mathrm{on}}/\mathrm{d}N$ using (\ref{eq:smallKstep}) gives 
\begin{equation}
\frac{\mathrm{d}N}{\mathrm{d}\omega_{\mathrm{on}}}=0 \quad \mathrm{on}\quad \frac{\omega_{\mathrm{on}}}{\omega_{\mathrm{on}}^\dag}=\frac{N-(\xi/(\gamma-1))}{2N+\xi}.
\end{equation}
Thus for the neutral curve to lie in $N>0$ requires
\begin{equation}
    N>\frac{\xi}{\gamma-1}\equiv \frac{\bar{\xi} v_0}{f_0(\gamma-1)},
\end{equation}
providing a lower bound on the number of linkers needed for oscillations in terms of the walking speed and stall force of a linker, and the drag on the spindle.

The period of oscillations along the neutral stability curve predicted using (\ref{eq:ODEfreq}) increases as $K$ decreases (Fig.~\ref{fig:PDE_N15}(b)); thus a reduction of restoring forces corresponds to longer periods of oscillation. The rapid increase of the period as ${\omega}_{\textrm{on}}\rightarrow{\omega}_{\textrm{on}}^\dag$ coincides with $N\rightarrow\infty$. (\ref{eq:ODEfreq}) is well matched with (\ref{eq:grillK}) determined by \cite{grill2005theory}, as well as with the periods along the approximate stability curve identified by numerical solutions of the PDEs. 

\subsection{The structure of relaxation oscillations}
\label{sec:relax}

A further simplification to the model can be implemented by exploiting $\hat{K}$ as a small parameter. The approximately linear sections of $\tilde{z}$ (\hbox{e.g.} Fig.~\ref{fig:PDEsol}(c)) scale like ${\hat{K}}^{-1}$ in both time and amplitude, and are interrupted by rapid changes in spindle direction. 
Re-scaling $\tilde{t}=\tilde{\tilde{t}}/\hat{K}$ and $\tilde{z}=\tilde{\tilde{z}}/\hat{K}$ such that $\tilde{z}_{\tilde{t}}=\tilde{\tilde{z}}_{\tilde{\tilde{t}}}$, the ODEs (\ref{eq:ODE}) describing the slower phases of the dynamics become
\begin{subequations}
\label{eq:smallKstep12}
\begin{align}\label{eq:smallKstep1}
    \left(1+\rho e^{\gamma\left(1\mp\tilde{\tilde{z}}_{\tilde{\tilde{t}}}\right)}\right)\hat{B}^{\pm} + \hat{K}\hat{B}^{\pm}_{\tilde{\tilde{t}}} &= 1, \\
    \left(\hat{\xi}+\hat{B}^++\hat{B}^-\right)\tilde{\tilde{z}}_{\tilde{\tilde{t}}}+\tilde{\tilde{z}}&=\hat{B}^+-\hat{B}^-. \label{eq:smallKstep2}
    \end{align}
\end{subequations}
Posing expansions $\hat{B}^\pm = \hat{B}_0^\pm + \hat{K}\hat{B}_1^\pm + ... $ and $\tilde{\tilde{z}} = \tilde{\tilde{z}}_0+\hat{K}\tilde{\tilde{z}}_1 + ... \nonumber $, to leading order (\ref{eq:smallKstep12}) becomes
\begin{subequations}
\begin{align}\label{eq:smallKB0}
    \left(1+\rho e^{\gamma\left(1\mp\tilde{\tilde{z}}_{0,\tilde{\tilde{t}}}\right)}\right)\hat{B}_0^\pm=1, \\
    \left(\hat{\xi}+\hat{B}_0^++\hat{B}_0^-\right)\tilde{\tilde{z}}_{0,\tilde{\tilde{t}}}+\tilde{\tilde{z}}_0=\hat{B}_0^+-\hat{B}_0^-. \label{eq:smallKz}
    \end{align}
\end{subequations}
We may rewrite (\ref{eq:smallKz}) as
\begin{equation}
    \tilde{\tilde{z}}_0=\hat{B}_0^+-\hat{B}_0^--\left(\hat{\xi}+\hat{B}_0^++\hat{B}_0^-\right)\tilde{\tilde{z}}_{0,\tilde{\tilde{t}}}\equiv G\left(\tilde{\tilde{z}}_{0,\tilde{\tilde{t}}}\right), \label{eq:branch}
\end{equation}
with $\hat{B}_0^{\pm}$ defined by (\ref{eq:smallKB0}). 


{\color{black}The displacement-velocity reationship} (\ref{eq:branch}) approximates the slow phases of  the limit cycle in ($\tilde{\tilde{z}}_0$, $\tilde{\tilde{z}}_{0,\tilde{\tilde{t}}}$)-space as $\hat{K}\rightarrow0$. Recalling that ${y}^\pm_{\textrm{c}}=1\mp{z}_{{t}}$, then following a parameter rescaling, (\ref{eq:branch}) can also be used to describe the limit cycle in (${z}_0$, ${y}^\pm_{\textrm{c}}$) (black curve in Fig.~\ref{fig:ODESol}(c-e)). The limit cycles obtained by solving the ODE and PDE systems with equivalent parameters are shown to closely match with this expected limit cycle (Fig.~\ref{fig:ODESol}(c-e)). These cycles show the fast phases of the relaxation oscillation as the spindle pole changes its direction of motion (the approximately vertical sections at the maximum and minimum values of $\tilde{\tilde{z}}$). The maximum amplitude of oscillation can be estimated by the roots of $G$, which can be determined by solving
\begin{equation}
    \frac{\textrm{d}G}{\textrm{d}\tilde{\tilde{z}}_{0,\tilde{\tilde{t}}}}=0.
\end{equation}
for roots $G_{\textrm{max}}$ and $G_{\textrm{min}}$. Then the amplitude of oscillation during relaxation oscillations can be estimated by 
\begin{equation}\label{eq:amp}
    \tilde{z}\approx (G_{\textrm{max}} -G_{\mathrm{min}}) /\hat{K}.
\end{equation}
Thus the amplitude of oscillation can be estimated from the ratio of pulling to pushing ($\hat{K}$), the effective drag ($\hat{\xi}$), the ratio of the unbinding to binding rates ($\rho$) and the tension sensitivity of unbinding ($\gamma$).

This approximation also illustrates how the tension-sensitivity of the cortical force generators, mediated by $\gamma$, is key for oscillations. Setting $\gamma=0$ in (\ref{eq:smallKB0}) uncouples the values of $\hat{B}^\pm_0$ from the spindle pole dynamics, thus $\hat{B}^+_0=\hat{B}^-_0$ and (\ref{eq:smallKz}) becomes
 $   \tilde{\tilde{z}}_0 = -\left(\hat{\xi}+2\left(1+\rho\right)^{-1}\right)\tilde{\tilde{z}}_{0,\tilde{\tilde{t}}}$,
giving a linear relationship between $\tilde{\tilde{z}}_0$ and $\tilde{\tilde{z}}_{0,\tilde{\tilde{t}}}$ and eradicating the limit cycle. Thus the coupling of the populations of bound force generators through the tension-sensitive unbinding rate is required for oscillations, as was shown by stability analysis of the ODE system (\ref{eq:omedag}).

{\color{black}
\subsection{Testing the accuracy of the ODE system}
\begin{figure}
\centering
\includegraphics[width=0.7\textwidth]{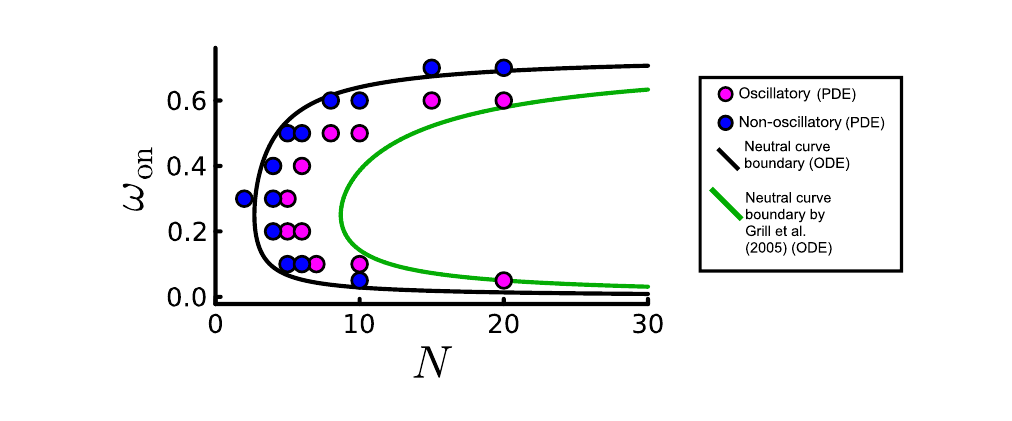}
\caption{{\color{black}{\textbf{Increasing $\omega_0$ results in a neutral curve which underestimates the threshold number of $N$.} Numerically solving the Fokker-Planck system (circles) reveals a boundary in $\left(N, \omega_{\textrm{on}}\right)$ space which separates oscillatory from non-oscillatory solutions. Each circle represents a numerical solution, labelled magenta if the spindle pole has sustained oscillations and blue if the spindle pole position decays to $z=0$ for large $t$. The black line represents the neutral curve separating oscillatory and decaying solutions as determined by stability analysis of the ODEs (\ref{eq:stab}) using equivalent parameters. The green curve shows the stability boundary (G1) predicted by  (\ref{eq:grillK}) from \cite{grill2005theory}. All parameters are as in Table~\ref{table:nondims} except that ${D_{\mathrm{b}}} = 8\times 10^{-3}$, ${D_{\mathrm{u}}}=4\times 10^{-3}$, and $\omega_0=0.1$.}}}
\label{fig:largeunbinding}
\end{figure}

For $\omega_0=0.001$, comfortably satisfying the condition $\omega_0\ll 1$ that allows the reduction of the Fokker--Planck system (\ref{eq:FPE}--\ref{eq:Pu+Pb=1}) to the ODEs (\ref{eq:ODE}), the latter make accurate predictions for the onset of sustained oscillations (Fig.~\ref{fig:PDE_N15}(a)).  In this limit, the stability threshold (\ref{eq:stab}) is almost indistinguishable from that of the heuristic model (\ref{eq:grillK}) proposed by \cite{grill2005theory}.  To test the robustness of each approximation, Fig.~\ref{fig:largeunbinding} shows predicted stability thresholds for $\omega_0=0.1$, against solutions of (\ref{eq:FPE}-\ref{eq:Pu+Pb=1}). A marked difference in the minimum value of $N$ required to cross the neutral curve is observed, with the threshold (\ref{eq:stab}) underestimating the lowest value of $N$ required to elicit oscillations in the PDE solutions (by 45\% for $\omega_{\mathrm{on}}=0.3$). Alternatively, (\ref{eq:grillK}) overestimates the threshold value of $N$ (by 77\%  for $\omega_{\mathrm{on}}=0.3$), demonstrating a modest benefit of the rigorously derived ODE system in this parameter regime.}

{\color{black}
\section{Characterising noise-induced oscillations}
\label{sec:noisosc}
}

We now use the ODE system (\ref{eq:ODE}) to provide further evidence that the oscillations in Fig.~\ref{fig:stochastic_ex}(a) are noise-induced.  Despite lying outside the neutral curve (Fig.~\ref{fig:PDE_N15}(a)), the period of the stochastic oscillations is well approximated by (\ref{eq:fullperiod}) (Fig.~\ref{fig:stochastic_ex}(a), red bar), indicating that noise due to the {\color{black} binding and unbinding of a} relatively small number of linkers {\color{black}may be} sufficient to overcome the damping evident in the Fokker--Planck description (Fig.~\ref{fig:PDE_N15}(c)) and in the ODE model.  As explained in Appendix~\ref{secA1}, the Fokker--Planck system (\ref{eq:FPE}, \ref{eq:poleeom}) proposed by \cite{grill2005theory} is a simplified form of the high-dimensional chemical Fokker--Planck equation associated with the full stochastic model; we attribute the failure of (\ref{eq:FPE}, \ref{eq:poleeom}) to predict the oscillations in Fig.~\ref{fig:stochastic_ex}(a) to this simplification, {\color{black} and show below how reintroducing stochastic effects associated with binding kinetics can explain some features of observations.}

{\color{black}
To do so, we adopt the framework outlined by \cite{boland2008} to estimate the amplitude of small-amplitude noise-induced oscillations.  At small amplitudes, it is appropriate to linearise the exponential term in (\ref{eq:ODE}) using $\mathrm{exp}(\pm\gamma \tilde{z}_{\tilde{t}})\approx 1\pm \gamma \tilde{z}_{\tilde{t}}$.  We then treat the ODE model (\ref{eq:ODE}) as a chemical kinetic system with eight reactions, written as
\begin{equation}
\left(
\begin{matrix}
    \hat{B}^+ \\ \hat{B}^- \\ \tilde{z}
\end{matrix}
\right)_{\tilde{t}}=
\left(
\begin{matrix}
1 & -1 & 0 & 1 & 0 & 0 & 0 & 0 \\
1 & 0 & -1 & 0 & -1 & 0 & 0 & 0 \\
0 & 0 & 0 & 0 & 0 &1 & -1 & -1
\end{matrix}
\right)
\left(
\begin{matrix}
    1 \\ \hat{B}^+ (1+\rho e^\gamma) \\ \hat{B}^- (1+\rho e^\gamma)\\ 
    \rho e^\gamma \hat{B}^+ a \\ 
    \rho e^\gamma \hat{B}^- a \\ 
    \hat{B}^+/(\hat{\xi}+\hat{B}^++\hat{B}^-) \\
    \hat{B}^-/(\hat{\xi}+\hat{B}^++\hat{B}^-) \\
    \hat{K}\tilde{z}/(\hat{\xi}+\hat{B}^++\hat{B}^-) 
\end{matrix}
\right),
\label{eq:d1}
\end{equation}
where $a\equiv \gamma(\hat{B}^+-\hat{B}^--\hat{K}\tilde{z})/(\hat{\xi}+\hat{B}^++\hat{B}^-)$.
The columns $\boldsymbol{\nu}_i$ of the $3\times 8$ stoichiometric matrix can be assembled with the reaction rates $a_i$ ($i=1,\dots,8$) to form the correlation matrix $\mathsf{D}=\tfrac{1}{2}\sum_i \boldsymbol{\nu}_i\boldsymbol{\nu}_i^\top a_i$, where 
\begin{equation}
    \mathsf{D}={{\frac{1}{2}}}\left(
    \begin{matrix}
    1+\hat{B}^+(\lambda+{{\rho e^\gamma}}a)  & 1 & 0 \\
    1 & 1+\hat{B}^-(\lambda +{{\rho e^\gamma}}a) & 0 \\
    0 & 0 & \frac{\hat{B}^++\hat{B}^-+\hat{K}\tilde{z}}{\xi+\hat{B}^++\hat{B}^-}
    \end{matrix}\right)
\end{equation}
and $\lambda\equiv 1+\rho e^\gamma$.  Evaluated at the equilibrium point (\ref{eq:Bsteady}), $\mathsf{D}$ simplifies to
\begin{equation}
    \mathsf{D}^*=\left(
    \begin{matrix}
    1& \frac{1}{2} & 0 \\    \frac{1}{2} & 1 & 0 \\
    0 & 0 & 1/\left(2+\hat{\xi}\lambda\right)
    \end{matrix}\right).
\end{equation}
Linearising (\ref{eq:d1}) about the equilibrium yields the Jacobian matrix $\mathsf{J}^*$ satisfying
\begin{equation}
    \mathsf{J}^*=\frac{1}{2+\lambda \hat{\xi}}\left(\begin{matrix} 
    (\lambda-1) \gamma -\lambda(2+\lambda\hat{\xi}) & -(\lambda-1) \gamma & - \hat{K} (\lambda-1) \gamma \\
    -(\lambda-1) \gamma & (\lambda-1) \gamma -\lambda(2+\lambda\hat{\xi}) &   \hat{K} (\lambda-1) \gamma \\   
    \lambda & -\lambda  & -\hat{K} \lambda  \end{matrix} \right).
\end{equation}
The eigenvalues of $\mathsf{J}^*$ satisfy (\ref{eq:cubic}).  For a particular set of parameters, including $\hat{K}_c$, $\hat{\xi}_c$ satisfying (\ref{eq:stab}), at which $\mathsf{J}^*=\mathsf{J}^*_c$ (say), the Jacobian has one real negative eigenvalue and a complex conjugate pair with zero real part and frequency $\Omega_c$ satisfying (\ref{eq:ODEfreq}).  Moving away from neutral stability by changing $N$ (\hbox{i.e.} moving horizontally in Fig.~5(a)) can be represented by setting $\hat{K}=\hat{K}_c(1-\epsilon)$, $\hat{\xi}=\hat{\xi}_c(1-\epsilon)$ for some $ \epsilon\equiv \delta N/N\ll 1$ (using (\ref{eq:odeparams})), with $\lambda$ and $\gamma$ remaining fixed.  Perturbing (\ref{eq:cubic}) in this way leads to complex eigenvalues
\begin{equation}
    s= \frac{\epsilon{\lambda}\left(\hat{K}_c+\lambda \hat{\xi}_c\right)}{2{\left(2+\lambda\hat{\xi}_c\right)}}\pm i\Omega_c\left(1-\frac{\epsilon}{2+\lambda\hat{\xi}_c}\right)+O(\epsilon^2),
    \label{eq:ev}
\end{equation}
confirming instability ($\mathrm{Re}(s)>0$) for an increase in $N$ ($\delta N>0$).  For $\epsilon<0$ (associated with a small reduction in $N$ from the neutrally stable case), we propose that the small negative growth rate (\ref{eq:ev}) balances the noise from stochastic forcing to determine the amplitude of noisy oscillations. 

Writing $\mathbf{x}(\tilde{t})=(\hat{B}^+,\hat{B}^-,\tilde{z})^\top$, the stochastic differential equation that generalises (\ref{eq:d1}) to describe small-amplitude noise-driven oscillations can then be written 
%
\begin{equation}
    \mathrm{d}\mathbf{x}=\mathsf{J}^*\mathbf{x}\,\mathrm{d}{\tilde{t}}+\mathbf{f}\,\mathrm{d}W \quad\mathrm{where}\quad \mathbf{f}\, \mathbf{f}^\top =2\mathsf{D}^*.
    \label{eq:sde}
\end{equation}
Here $\mathbf{f}=\sum_i \boldsymbol{\nu}_i \sqrt{a_i}$ and $W(\tilde{t})$ is a Wiener process.  Following \cite{gardiner1985}, when $\delta N<0$ (so that the eigenvalues of $\mathsf{J}^*$ have negative real part), the stationary covariance $\boldsymbol{\sigma}\equiv \langle \mathbf{x}(t),\mathbf{x}^\top(t)\rangle$ satisfies the Lyapunov equation $\mathsf{J}^*\boldsymbol{\sigma}+\boldsymbol{\sigma}\mathsf{J}^{*\top}=-2\mathsf{D}^*$.  (Equivalently, the stationary distribution of the Fokker--Planck equation associated with (\ref{eq:sde}) is proportional to $\mathrm{exp}[-\tfrac{1}{2} \mathbf{x}^\top \boldsymbol{\sigma}^{-1}\mathbf{x}]$.)  Writing 
\begin{equation}
\boldsymbol{\sigma}=
\left(\begin{matrix}a & b & c \\ b & d & e \\ c & e & f\end{matrix}\right),
\end{equation}
the coefficients satisfy
\begin{equation}\label{eq:lyap_f}
{\small{
\left(\begin{matrix}
\alpha &  -(\lambda-1)\gamma & -\hat{K}(\lambda-1)\gamma & 0 & 0 & 0\\ 
-(\lambda-1)\gamma & 2\alpha & \hat{K}(\lambda-1)\gamma &-(\lambda-1)\gamma & -\hat{K}(\lambda-1)\gamma &0\\ 
\lambda & -\lambda & \alpha-\hat{K}\lambda & 0&-(\lambda-1)\gamma & -\hat{K}(\lambda-1)\gamma \\
0 & -(\lambda-1)\gamma & 0 & \alpha & \hat{K}(\lambda-1)\gamma & 0\\ 
0 & \lambda & -(\lambda-1)\gamma &-\lambda &\alpha-\hat{K}\lambda &\hat{K}(\lambda-1)\gamma\\ 
0 & 0 & \lambda & 0 & -\lambda & -\hat{K}\lambda 
\end{matrix}\right)
\left(\begin{matrix}a \\ b \\ c \\ d \\ e \\ f\end{matrix}\right)=
\left(\begin{matrix}-(2+\lambda \hat{\xi}) \\ -(2+\lambda \hat{\xi}) \\ 0  \\ -(2+\lambda \hat{\xi}) \\ 0 \\ -1 \end{matrix}\right)
}}
\end{equation}
where $\alpha\equiv (\lambda-1)\gamma-\lambda(2+\lambda\hat{\xi})$.  We can use $\sqrt{f}$ to estimate the amplitude of noise-driven oscillations in Fig.~3(a) lying outside the neutral curve. Solving (\ref{eq:lyap_f}) for $f$ gives
\begin{equation} \label{eq:f}
    f = \frac{4\gamma\alpha\left(\lambda-1\right)+\lambda^2\left(2+\lambda\hat{\xi}\right)\left(3+\lambda\hat{\xi}+\hat{K}\right)}{\hat{K}\lambda^2\left(2+\lambda\hat{\xi}\right)\left(\lambda\left(2+\lambda\hat{\xi}+\hat{K}-2\gamma\right)+2\gamma\right)}.
\end{equation}
Perturbing about the neutral curve, $f$ simplifies to
\begin{equation}
    f\approx \frac{2\gamma(\lambda-1)\hat{K}_c+\lambda (2+\lambda\hat{\xi}_c)}{2 \hat{K}_c \lambda(2+\lambda\hat{\xi}_c)^2 \left[-\Re(s)\right]}
\label{eq:famp}
\end{equation}
in terms of the eigenvalues (\ref{eq:ev}).  Clearly this estimate of $f$ is unbounded as $\vert\Re(s)\vert\rightarrow 0$, violating the small-amplitude assumption and suggesting that (\ref{eq:famp}) is best considered as an approximate upper-bound of the true amplitude.

Using the parameters for the stochastic simulation in Fig.~\ref{fig:stochastic_ex}(a) (a further example is shown in Fig.~\ref{fig:D63}(c)), $\sqrt{f}/\omega_{\textrm{on}}\approx 510$.  Approaching the neutral curve by increasing $N$ from 15 to 18 (Fig.~\ref{fig:D63}(d)) increases $\sqrt{f}/\omega_{\textrm{on}}$ by 35\% to approximately 690.  Figs~\ref{fig:D63}(e,f) show that the interquartile range of the density of $z$ values of stochastic oscillations increase by 38\% (from 59.5 when $N=15$ to 82.1 when $N=18$).  $\sqrt{f}$ therefore captures the trend in amplitude but overestimates its magnitude by approximately a factor of 8.  Discrepanices may arise from a number of sources, including linearisation of the exponential term in (\ref{eq:ODE}), linearisation leading to (\ref{eq:famp}), interaction with random motion of force generators and insufficient sampling of stochastic time series.


\begin{figure}
\includegraphics[width=\textwidth]{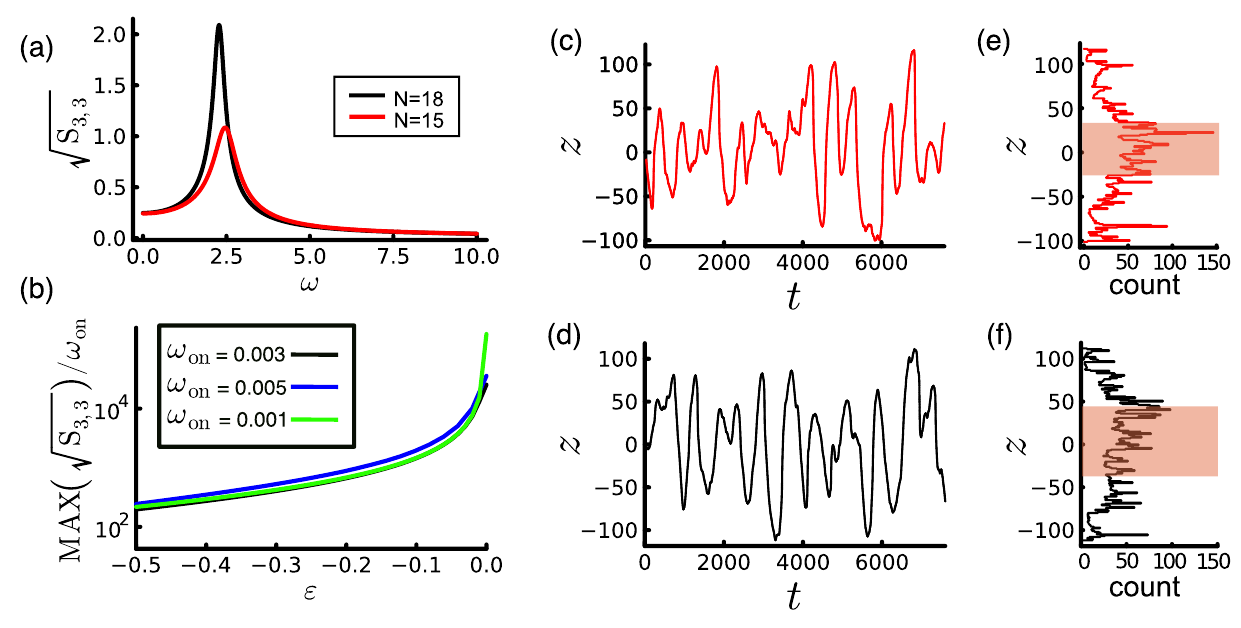}
\caption{{\color{black}{\textbf{Amplitude estimation of noise-induced oscillations.} (a) The $(3,3)$ component of spectrum matrix $\mathsf{S}$ using (\ref{eq:S}), for parameters as in Table \ref{table:nondims} with $N=18$ (black) and $N=15$ (red). (b) Amplitude estimation from $\mathsf{S}$ versus distance to the neutral curve $\epsilon=\delta N/N$ for $\omega_{\textrm{on}}=0.003$ (black), $\omega_{\textrm{on}}=0.005$ (blue) and $\omega_{\textrm{on}}=0.001$ (green). (c-d) Example pole dynamics from stochastic simulations and (e-f) the corresponding histograms weighted by time spent at each $z$-position. Shaded regions denote the interquartile range. Parameters as in Table \ref{table:nondims} with (c,e) $N=15$, (d,f) $N=18$.}}}
\label{fig:D63}
\end{figure}

The spectrum matrix $\mathsf{S}$ (the Fourier transform of the time correlation matrix in the stationary state) satisfies \citep{gardiner1985}
\begin{equation}\label{eq:S}
\mathsf{S}(\omega) 
=\frac{1}{\pi}(i\omega-\mathsf{J}^*)^{-1}\mathsf{D}^*(-i\omega-\mathsf{J}^*)^{-\top}.
\end{equation}
The $\tilde{z}\tilde{z}$ component of $\mathsf{S}$ is plotted in Fig.~\ref{fig:D63}(a) using the parameters for the stochastic simulation in Fig.~3(a) ($N=15$), and for $N=18$ (approaching the neutral curve).  Equation (\ref{eq:S}) predicts a sharpening of the spectrum with an approximate doubling of the root-mean-square amplitude  near the resonant frequency; the maximum of $\sqrt{\mathsf{S}_{3,3}}/\omega_{\textrm{on}}$ corresponds to $z\approx360$ and $z\approx700$ for $N=15$ and $N=18$ respectively, broadly consistent with values of $\sqrt{f}/\omega_{\textrm{on}}$.  The relation $\boldsymbol{\sigma}=\int_{-\infty}^\infty \mathsf{S}(\omega)\,\mathrm{d}\omega$ emphasises the contributions of a narrower range of frequencies to the noise-induced oscillation as $N$ increases.  Accordingly, Figs~\ref{fig:D63}(c,d) show a more coherent oscillation for larger $N$.  Fig.~\ref{fig:D63}(b) illustrates similar increases in the predicted amplitude of stochastic oscillation as the neutral curve in Fig.~\ref{fig:PDE_N15}(a) is approached along different values of $\omega_{\mathrm{on}}$; again this is best interpreted as a likely upper bound. 
}
\bigskip

\section{Discussion}\label{sec12}

We have investigated the factors promoting relaxation and noise-driven oscillations of the mitotic spindle identified experimentally (Fig.~\ref{fig:example}), by revisiting the mathematical model proposed by \cite{grill2005theory}. To this end, we used stochastic simulations to demonstrate the effect of noise on 1D pole movement (Section~\ref{Gillespie}, Fig.~\ref{fig:stochastic_ex}); this involved $2N$ random walkers (linkers) switching between bound and unbound states, with their motion coupled via an ODE to that of the spindle. The corresponding {\color{black}mean-field} Fokker--Planck equations (Section~\ref{sec:FP}, Fig.~\ref{fig:PDEsol}), involving four PDEs coupled to an ODE, were reduced systematically (Section~\ref{sec:odes}) to three nonlinear ODEs (\ref{eq:ODE}).  {\color{black}When binding kinetics is slow ($\omega_0\ll 1$),} these turn out to be a special (and simpler) case of the ODE system presented by \cite{grill2005theory}, and {\color{black}both systems} show close agreement with the Fokker--Planck solutions in predicting conditions necessary for the onset of self-excited oscillations (Fig.~\ref{fig:PDE_N15}(a)).  {\color{black}(The ODE systems deviate as $\omega_0$ increases, with (\ref{eq:ODE}) being marginally more accurate than the Grill \hbox{et al.} model for $\omega_0=0.1$ (Fig.~\ref{fig:largeunbinding}).)}  Further asymptotic reduction of (\ref{eq:ODE}) revealed the single algebraic equation describing the slow dynamics of the nonlinear relaxation oscillations and the associated amplitude of oscillation (Section~\ref{sec:relax}, Fig.~\ref{fig:ODESol}(c-e)).  

While there is consistency between the descriptions in many respects, a striking feature is the appearance in stochastic simulations of noise-induced oscillations in a regime predicted to be linearly stable by the {\color{black}mean-field} Fokker--Planck description (and the associated ODE system).  The oscillations arise close to a stability boundary, and their period is well predicted by analysing the three linearised ODEs (the green circle in Fig.~\ref{fig:PDE_N15}(a) highlights the position in parameter space occupied by the stochastic solution in Fig.~\ref{fig:stochastic_ex}(a); the period prediction is given by (\ref{eq:fullperiod})).  {\color{black}By restoring a representation of stochastic binding/unbinding kinetics (Section \ref{sec:noisosc}), we provide evidence that} the amplitude of the oscillations is likely regulated by the noise associated with {\color{black}binding kinetics; the approximate SDE (\ref{eq:sde}) captures amplitudes to within an order of magnitude (Fig.~\ref{fig:D63})}.  
{\color{black}  The relationship between cell shape and division orientation was first described in the 1880s \citep{hertwig1884} but it is inherently noisy (e.g. \cite{nestor2019decoupling, lam2020isotropic, bosveld2016epithelial}). This makes trying to understand more complex regulation of spindle orientation, such as regulation by external forces, very challenging and can require experimental analysis of many spindles to obtain meaningful results.  Our study reveals two underlying mechanisms of oscillation: small-amplitude spindle oscillations driven by noisy binding kinetics; and the previously described  larger-amplitude oscillations driven by noise in force generator motion.  Together, these may explain some of the noise seen in spindle and division orientation.  For example,}  
the predicted period of {\color{black}experimentally-observed} oscillation can be used to infer (or at least constrain estimates of) some of the parameters relevant to the \emph{Xenopus} system {\color{black}illustated in Fig.~\ref{fig:example}}. For parameters as in Table~\ref{table:params}, for a period of $T\approx100$~s as seen experimentally (Fig.~\ref{fig:example}(d)) then $N=175$ force generators would be required. However, if restoring forces were reduced to $K=0.005$ then $N=21$ force generators would be required to achieve the same period. These would result in oscillations with an amplitude of $\approx15~\mu$m which is comparable with the typical size of a cell ($\approx$20~$\mu$m diameter) though approximately three to four times larger than what was recorded experimentally (Fig.~\ref{fig:example}(d), $\approx2$-$5~\mu$m). {\color{black} Further work is needed to refine assessments of parameters to allow more direct comparison between theory and experiment.}

Fig.~\ref{fig:PDE_N15}(c,d) shows how increasing demographic stochasticity by increasing the value of diffusive terms $D_{\textrm{b}}$ and $D_{\textrm{u}}$ can promote oscillations.  Thus, noise {\color{black}associated with movement of force generators} increases the ease with which oscillations are sustained.  Expansion of regions of parameter space giving rise to oscillatory solutions under the addition of noise has been seen elsewhere, in studies of oscillations in protein expression.  For example, \cite{phillips2016stochasticity} show that stochasticity increases the robustness of the oscillatory phenotype of gene expression resulting in the correct timing of cell differentiation.  Indeed, if oscillations of the spindle pole play an important role in correctly orientating the mitotic spindle, then the inherent stochasticity of biological systems due to fluctuations in protein levels or ATP availability (driving dynein movement) would aid the robustness of correct spindle orientation.

Relaxation oscillations arise when restoring forces decrease relative to pulling forces (Fig.~\ref{fig:ODESol}(c-e)). Biologically, this may be achieved by reducing the restoring force, for example by hinging of microtubules at the spindle pole \citep{howard2006elastic, rubinstein2009elasticity}, or by increased numbers of dynein linkers at the cell cortex (denoted in this work by an increase in $N$). The resulting linear sections of the spindle pole oscillation align temporally with slow phases in the time-evolution of $P_{\textrm{b}}^\pm$ and $y_{\textrm{c}}^\pm$ (Fig.~\ref{fig:ODESol}(c-e)), until the spindle pole is sufficiently displaced from the centre for the restoring force to create a rapid reversal in the spindle pole velocity $z_{t}$. This change in the spindle pole velocity results in a rapid increase in the value of $y_{\textrm{c}}$ on the opposite cortex, which in turn creates a rapid decrease in the value of $P_{\textrm{b}}^\pm$ due to the tension-sensitivity of the unbinding rate. $P_{\textrm{b}}^\pm$ and $y_{\textrm{c}}^\pm$ have amplitudes that are self-limited by the tension-sensitive unbinding, and amplified by their connection to the motion of the spindle pole. We have shown that $\gamma>1$ is required for oscillations to occur at all (e.g in (\ref{eq:omedag}), (\ref{eq:lower})); dynein's slip--bond \citep{ezber2020dynein} is crucial for oscillatory dynamics, but the sensitivity of the slip--bond to tension may affect the nonlinearity of the oscillation. 

In the limit of small restoring forces, the model may be simplified to a single algebraic equation describing the slow phases of the limit cycle in ($\tilde{\tilde{z}}_0, \tilde{\tilde{z}}_{0,\tilde{\tilde{t}}}$) and subsequently in ($z_0,y_{\textrm{c}}^\pm$), where the maximum amplitude of oscillation can be estimated using (\ref{eq:amp}).  {\color{black}This relation has similarities with the bistable force-velocity relation derived by \cite{schwietert2020bistability} in their model of kinetochore-chromosome dynamics, which underlies relaxation oscillations in that system.}. Re-dimensionalisation of the amplitude seen during relaxation oscillations (Fig.~\ref{fig:ODESol}(c-e)) results in an oscillation with an amplitude of order $\approx0.1$~mm (for chosen baseline parameters), which is an order of 10 larger than the typical size of a cell in the \textit{Xenopus} epithelium ($\approx$20~$\mu$m diameter). {\color{black}This large amplitude is an artefact of the linear force-displacement law (\ref{eq:polepos}) and the 1D description;  the imposed geometry necessary for a 2D description would allow the redistribution of pulling forces away from the direction of motion upon close proximity of the spindle pole as in \cite{wu2024laser}. Indeed, without pushing forces a redistribution of pulling forces is sufficient to cause reversal of spindle motion and relaxation oscillations \citep{wu2024laser}. This effect could be described by a nonlinear restoring force at the boundary in 1D.  }

While we have studied a 1D model, imaged mitotic spindles in epithelial cells show 2D dynamics (Fig.~\ref{fig:example}; Online Resource 1), with forces acting on both spindle poles originating from the entire cell periphery. To properly consider the spindle dynamics and infer \emph{Xenopus} system specific parameters, the relative motion of the two spindle poles must become part of the equation. The simplification of the present model to ODEs or low-dimensional SDEs is a key step to fully modelling 2D movements of a full mitotic spindle. From there, we may begin to piece together the full processes by which the mitotic spindle is positioned and orientated in tissue-based cells. In doing so, we must remain mindful that inherent stochasticity may increase the mobility of the spindle.

\backmatter

\bmhead{Supplementary information}

Online Resource 1: Time--lapse of a dividing cell in the epithelium of a $\emph{Xenopus laevis}$ embryo at stage 10. The mitotic spindle is seen in green (GFP-$\alpha$-tubulin) and metaphase plate in magenta (mCherry-Histone 2B). Images taken every $t=6.0~$s. 

Online Resource 2: a) Movie of the evolution of fluxes $J^+_{\textrm{b}}$, $J^+_{\textrm{u}}$ and their sum $J^+_{\textrm{b}}+J^+_{\textrm{u}}$. b) as in a) with a truncated $y$-axis to better demonstrate the dynamics of $J^+_{\textrm{b}}$ and $J^+_{\textrm{b}}+J^+_{\textrm{u}}$. c) The evolution of the pdfs $P^+_{\textrm{u}}$ and $P^+_{\textrm{b}}$. Parameter values as in Table~{\ref{table:params}} but with $D_{\textrm{b}}=0.008$, $D_{\textrm{u}}=0.004$ and $N=25$.
 
\bmhead{Acknowledgments}

The authors have applied a Creative Commons Attribution (CCBY) licence to any Author Accepted Manuscript version arising.  This work was supported by the Wellcome Trust (098390/Z/12/Z and 225408/Z/22/Z). DH was supported by a Wellcome Trust PhD studentship (220054/Z/19/Z).  OEJ and SW acknowledge support from the Leverhulme Trust (RPG-2021-394).  The authors are grateful to Simon Cotter for technical advice {\color{black}and to anonymous reviewers for helpful suggestions}.

\section*{Declarations}

\bmhead{Author contributions}

All authors contributed to the study conception and design. Mitotic spindle movie acquisition was performed by SW. Imaging of the metaphase plate and data analysis was performed by DH. The first draft of the manuscript was written by DH and all authors commented on subsequent versions of the manuscript. All authors read and approved the final manuscript.

\bmhead{Ethical approval}
 All work with \textit{Xenopus laevis} was performed using protocols approved by the UK Government Home Office under the Home Office Project Licence PFDA14F2D (Holder: Professor Enrique Amaya) and Home Office Personal Licences held by SW and DH.
\bmhead{Availability of data and materials}
Biological data available upon request.
\bmhead{Code availability}
Code available as follows: Stochastic simulations (Section \ref{Gillespie}) at \url{github.com/dionn-hargreaves/StochasticSimulation_SpindleMovements}. Fokker-Planck equations (Section \ref{sec:FP}) at \url{github.com/dionn-hargreaves/1DSpindle_FP_MoL}. ODE equations (Section \ref{sec:odes}) at \url{github.com/dionn-hargreaves/ODE_1D_spindle}.

\begin{appendices}



\section{Acquisition of biological data}\label{sec:A3}

\textbf{\textit{Xenopus laevis}}
\textit{Xenopus laevis} male and female frogs were housed within tanks maintained by the in-house animal facility at the University of Manchester.  Female frogs were pre-primed 4-7 days in advance of embryo collection by injection with 50 U of pregnant mare serum gonadotrophin (Intervet UK) into the dorsal lymph sac. One day prior to embryo collection, male and pre-primed female frogs were primed by injection with 100 U (male) and 200 U (female) of human chorionic gonadotrophin (hCG; Chorulon, MSD) into the dorsal lymph sac. 2-5 h ahead of embryo collection, primed male and female frogs were transferred into the same tank for amplexus. Embryos were collected over 1 h time periods. Embryos were dejellied with 2\% L-cysteine solution (Sigma Aldrich, \#168149-100G) in 0.1X Marks Modified Ringers (MMR) [1X MMR: 100mM NaCl, 2mM KCl, 1mM MgCl2, 1mM CaCl2, 0.5mM EDTA and 5mM HEPES, pH7.8],  rinsed with 0.1X MMR and incubated at room temperature (RT) to reach two-cell stage.

All \textit{Xenopus} work was performed using protocols approved by the UK Government Home Office and covered by Home Office Project License PFDA14F2D (License Holder: Professor Enrique Amaya) and Home Office Personal Licenses held by SW and DH.

\subsection{Whole embryo movies, spindle and metaphase plate}
For live imaging of mitotic spindles in \textit{Xenopus} embryos (Fig.~\ref{fig:example}(a, b)), both cells of two-cell embryos were microinjected with $5$~nl of mRNA for GFP–$\alpha$-tubulin (needle concentration of 0.5~g/l) and mCherry–Histone 2B (0.1~g/l), to highlight spindle microtubules and chromosomes, respectively. Embryos were incubated for 20 hours post fertilization at 16$^\circ$C and then mounted for live imaging in 0.1X Marks Modified Ringers (MMR) [10X solution: 1~M NaCl, 20~mM KCl, 10~mM MgCl$_2$.6H$_2$O, 20~mM CaCl$_2$.2H$_2$O, 1~mM EDTA disodium salt, 50~mM Hepes, up to 5~L with distilled water], using a ring of vacuum grease to contain the embryos and support a glass coverslip as in \cite{woolner2010imaging}. Imaging took place at developmental stages 10–11. Single focal plane live-cell images of spindles were collected at RT (21$^\circ$C) every 6~seconds using a confocal microscope (FluoView FV1000; Olympus) with FluoView acquisition software (Olympus) and a 60X, 1.35~NA U Plan S Apochromat objective. Time-lapse videos were constructed from the single focal plane images using ImageJ.

\subsection{Animal cap movies, metaphase plate only}
For imaging in the animal cap to follow dynamics of spindle movements (Fig.~\ref{fig:example}(c-f)), both cells of two-cell embryos were microinjeceted with mRNA for mCherry-Histone 2B (0.1~g/l; to highlight chromosomes) and BFP-CAAX (0.1~g/l; to highlight cell edges) using a Picospritzer III Intracel injector (\emph{Parker instrumentation}). Injected embryos were washed in 0.1\% MMR and incubated in fresh 0.1\% MMR overnight at $16^\circ$C. Animal cap explants were prepared from the injected embryos at the early gastrula stage (stage 10). The embryos were transferred to Danilchik's for Amy explant culture media (DFA) [53~mM NaCl, 5~mM Na$_2$CO$_3$, 4.5~mM Potassium gluconate, 32~mM Sodium gluconate, 1~mM CaCl$_2$, 1~mM MgSO$_4$, up to 1~L with MilliQ water, pH 8.3 with Bicine] in 0.1\% BSA (\emph{Sigma, A7906}). The vitelline membranes were removed from the embryos using forceps, and the explant removed by incisions with the forceps around the animal pole resulting in separation of the animal cap tissue from the embryo~\citep{joshi2010live}. The animal caps were then transferred onto a fibronectin-coated PDMS membrane with the basal side in contact with the membrane to prevent balling up and held in place with a coverslip. The caps were then left to recover for 2 hours at $18^\circ$C before imaging \citep{goddard2020applying}.

The PDMS membranes were prepared as described previously \citep{goddard2020applying}. The PDMS membrane was mounted onto a stretch apparatus (\emph{custom made by Deben UK 722 Limited}) and subjected to a uniaxial stretch of 0.5~mm displacement to ensure that the membrane remained taut under gravity and the weight of the animal cap~\citep{nestor2018relating}. Images were acquired every 5~seconds on a Leica TCS SP8 AOBS upright confocal using a 20X dipping objective at 2X confocal zoom. The confocal settings were as follows: pinhole 1.9 airy unit, 600~Hz bidirectional scanning, format 1024~x~1024. Images were collected using hybrid detectors with the detection mirror settings for red and blue at 600-690~nm, and 415-516~nm respectively, using the white light laser with excitation at 586~nm, and 405~nm laser lines. The images were collected non-sequentially with a $z$-spacing of 10~${\mu}$m between sections. Images were taken continuously without resetting for drifting in order to ensure no missing data points for dividing cells. The maximum imaging duration per animal cap was 2 hours. 

\subsection{Spindle movement analysis}
All images were processed and analysed on \texttt{ImageJ} \citep{schneider2012nih}. All measurements were taken from maximum intensity projections of the $z$-stack images. Each end of the metaphase plate was tracked in each frame using the \texttt{ImageJ} multi-point tool, returning an $x$ and $y$ coordinate for each point. The centre of the cell at the beginning of metaphase, $\mathbf{R}_1$, and the end of metaphase, $\mathbf{R}_2$, were used to create a linear correction to the metaphase plate position across metaphase time. Cell edges and tricellular vertices were manually traced at the beginning and end of metaphase using the \texttt{ImageJ} `Paintbrush tool' (brush width = 1 pixel). The manual traces were processed using in-house \texttt{Python} scripts to return the cell centre position \citep{nestor2018relating, nestor2019decoupling}. The measurements were made based on the polygonised cell according to the positions of the tricellular vertices. Oscillations were detected from signals using a periodogram. 

\section{Implementation of the Gillespie algorithm}\label{secA1}

The extension of an elastic linker is discretised into states $y^{\left(n\right)\pm,i}_{\textrm{b(u)}}$ with $i=0,1\ldots M$, separated by a fixed distance $\Delta {y}$ such that $y^{\left(n\right)\pm,i+1}_{\textrm{b(u)}} = y^{\left(n\right)\pm,i}_{\textrm{b(u)}} + \Delta{y}$. Each force generator $n$ has identifiers which denote the associated cortex ($\pm$), the current extension state ($i$), and the binding state ($\textrm{u}$ for unbound, $\textrm{b}$ for bound). The binding state will be identified in the subscript and written as $\textrm{b(u)}$, referring to a subscript b or u.

\begin{figure}
\includegraphics[width=\textwidth]{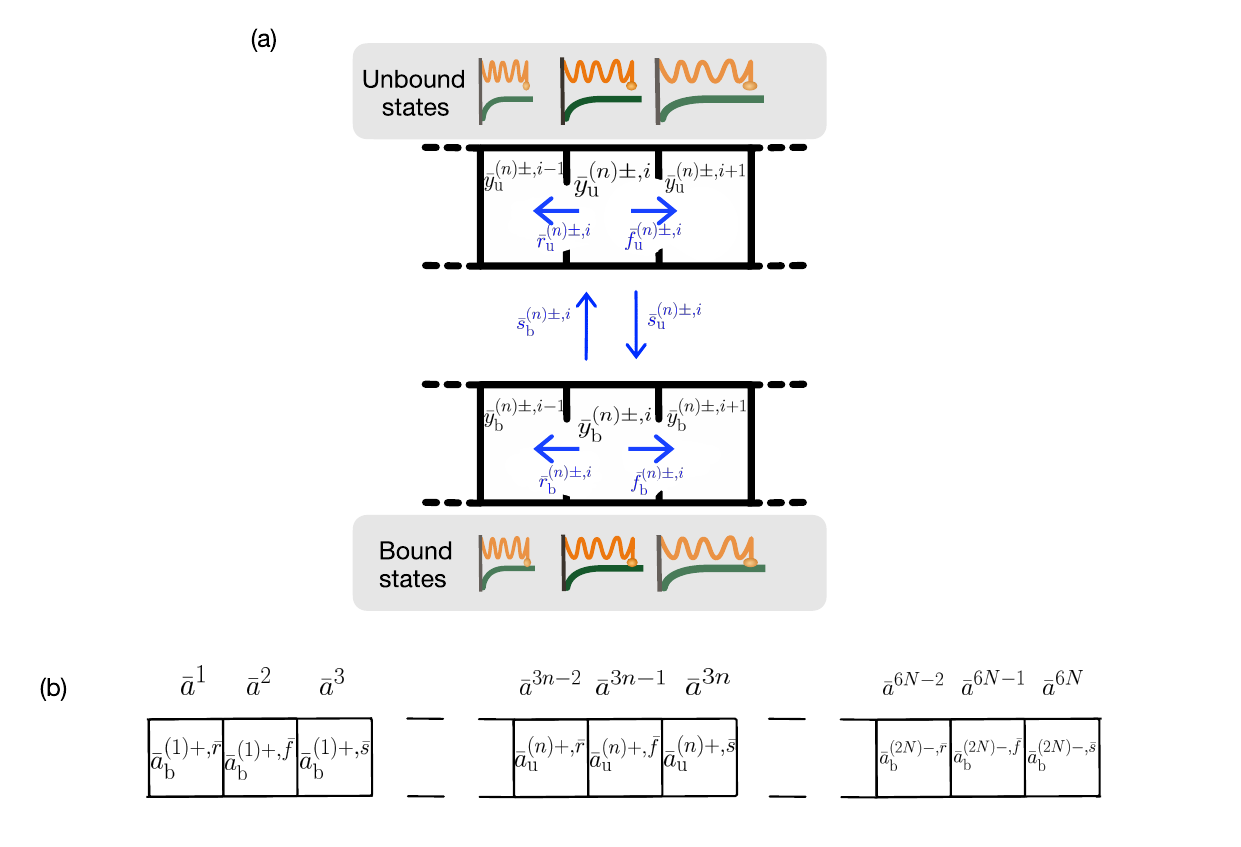}
\caption{\textbf{Graphical map of extension states for unbound and bound force generators.} (a) Unbound generators in state ${y}^{\left(n\right)\pm,i}_\textrm{u}$ may extend or retract with probabilities ${f}^{\left(n\right)\pm,i}_\textrm{u}$ and ${r}^{\left(n\right)\pm,i}_\textrm{u}$. Bound generators in state ${y}^{\left(n\right)\pm,i}_\textrm{b}$ may extend or retract with probabilities ${f}^{\left(n\right)\pm,i}_\textrm{b}$ and ${r}^{\left(n\right)\pm,i}_\textrm{b}$. Bound generators may unbind or vice-versa with rate constants ${s}^{\left(n\right)\pm,i}_\textrm{b}$ and ${s}^{\left(n\right)\pm,i}_\textrm{u}$ respectively. Diagrams of force generators show corresponding extension and binding states. Each individual force generator $n$ exists within these states. (b) Concatenated list of rate triplets to show numbering scheme. Probabilities from ${a}^1$ to ${a}^{3N}$ correspond to force generators $1\rightarrow N$ which exist in the upper cortex. Probabilities ${a}^{3N+1}$ to ${a}^{6N}$ correspond to force generators $N+1\rightarrow 2N$ which exist in the lower cortex.}
\label{fig:stochastic_states}
\end{figure}

At any time, a generator may
\begin{itemize}
    \item retract: ${y}^{\left(n\right)\pm,i}_{\textrm{b(u)}}\rightarrow{y}^{\left(n\right)\pm,i-1}_{\textrm{b(u)}}$ with probability $r^{\left(n\right)\pm,i}_{\textrm{b(u)}}$;
    \item extend: $y^{\left(n\right)\pm,i}_{\textrm{b(u)}}\rightarrow{y}^{\left(n\right)\pm,i+1}_{\textrm{b(u)}}$ with probability $f^{\left(n\right)\pm,i}_{\textrm{b(u)}}$; or
    \item switch between bound and unbound states: ${y}^{\left(n\right)\pm,i}_{\textrm{b}}\leftrightarrow {y}^{\left(n\right)\pm,i}_{\textrm{u}}$ with probability $s^{\left(n\right)\pm,i}_{\textrm{b(u)}}$.
\end{itemize}
These state-changing events are illustrated graphically in Fig.~\ref{fig:stochastic_states}(a). 

Probabilities $r^{\left(n\right)\pm,i}_{\textrm{b(u)}}$, $f^{\left(n\right)\pm,i}_{\textrm{b(u)}}$, and  $s^{\left(n\right)\pm,i}_{\textrm{b(u)}}$ are related to model parameters as follows. The switching probabilities were chosen such that an unbound generator may switch to become a bound generator within a short time $\tau$ with a probability $s^{\left(n\right)\pm,i}_{\textrm{u}}=\tau\omega_{\textrm{on}}$ for a constant binding rate $\omega_{\textrm{on}}$. 
In a short time $\tau$, a bound generator may unbind with probability $s^{\left(n\right)\pm,i}_{\textrm{b}}=\tau\omega_0e^{\gamma y^{\left(n\right)\pm,i}_{\textrm{b}}}$. 

In order to obtain expressions for $r^{\left(n\right)\pm,i}_{\textrm{b(u)}}$ and $f^{\left(n\right)\pm,i}_{\textrm{b(u)}}$,   consider
\begin{subequations}
    \begin{align}
\label{eq:velocity_probs}
{v^{\left(n\right)\pm,i}_{\textrm{b}(u)}}&=\frac{\Delta{y}}{\tau}\left(f^{\left(n\right)\pm,i}_{\textrm{b(u)}}-r^{\left(n\right)\pm,i}_{\textrm{b(u)}}\right), \\
 \label{eq:diffusion_probs}
{D^{\left(n\right)\pm}_{\textrm{b}(u)}}&=\frac{\left(\Delta{y}\right)^2}{2\tau}\left(f^{\left(n\right)\pm,i}_{\textrm{b(u)}}+r^{\left(n\right)\pm,i}_{\textrm{b(u)}}\right),
 \end{align} 
 \end{subequations}
as an effective drift speed and diffusion coefficient for force generators respectively. These arise from considering extension or contraction of each linker as a biased random walk. 
We summarise these transition probabilities as
\begin{subequations}
            \label{eq:trprob}
\begin{align}
    s^{\left(n\right)\pm,i}_{\textrm{b}}&={\tau}{\omega}_0e^{\gamma{y}^{\left(n\right)\pm,i}_{\textrm{b}}},  &s^{\left(n\right)\pm,i}_{\textrm{u}}&={\tau}{\omega}_{\textrm{on}},  \\
         r^{\left(n\right)\pm,i}_{\textrm{b}}&={\tau}\left(\frac{{D_{\mathrm{b}}}}{\left(\Delta{y}\right)^2}-\frac{{v}^{(n)\pm,i}_{\textrm{b}}}{2\Delta{y}}\right), \quad &r^{\left(n\right)\pm,i}_{\textrm{u}}&={\tau}\Gamma\left(\frac{{D_{\mathrm{u}}}}{\left(\Delta{y}\right)^2}+\frac{{y}^{(n)\pm,i}_{\textrm{u}}}{2\Delta{y}}\right),  \\
            f^{\left(n\right)\pm,i}_{\textrm{b}}&={\tau}\left(\frac{{D_{\mathrm{b}}}}{\left(\Delta{y}\right)^2}+\frac{{v}^{(n)\pm,i}_{\textrm{b}}}{2\Delta{y}}\right),  \quad &f^{\left(n\right)\pm,i}_{\textrm{u}}&={\tau}\Gamma\left(\frac{{D_{\mathrm{u}}}}{\left(\Delta{y}\right)^2}-\frac{{y}^{(n)\pm,i}_{\textrm{u}}}{2\Delta{y}}\right).
\end{align}
\end{subequations}
No flux conditions were enforced by setting $r^{(n)\pm,i=0}_{\textrm{b(u)}}=0$ and $f^{(n)\pm,i=M}_{\textrm{b(u)}}=0$.

The Gillespie algorithm \citep{gillespie1977exact} stipulates that the probability of a state-changing event (extension, retraction, or switch) happening within a short time $\tau$ is exponentially distributed with rates $r_{\textrm{b(u)}}^{\left(n\right)\pm,i}/\tau$, $f_{\textrm{b(u)}}^{\left(n\right)\pm,i}/\tau$, and  $s_{\textrm{b(u)}}^{\left(n\right)\pm,i}/\tau$, which sum together to give a total rate
\begin{equation}
R=\frac1\tau{{\sum}}_{n=1}^{2N}\left(r_{\textrm{b(u)}}^{(n)\pm,i}+f_{\textrm{b(u)}}^{(n)\pm,i}+s_{\textrm{b(u)}}^{(n)\pm,i}\right) .
\end{equation}
Here $2N$ is the total number of force generators within the system ($N$ per cortex), each of which is associated with either the upper (+) or lower (-) cortex, has an extension state $i$, and is either bound (b) or unbound (u). We assume that only one event for one force generator may occur, removing the possibility of simultaneous events. As $r_{\textrm{b(u)}}^{\left(n\right)\pm,i}$, $f_{\textrm{b(u)}}^{\left(n\right)\pm,i}$ and  $s_{\textrm{b(u)}}^{\left(n\right)\pm,i}$ are proportional to the short time $\tau$ (\ref{eq:trprob}),
the rates $r_{\textrm{b(u)}}^{\left(n\right)\pm,i}/\tau$, $f_{\textrm{b(u)}}^{\left(n\right)\pm,i}/\tau$, and  $s_{\textrm{b(u)}}^{\left(n\right)\pm,i}/\tau$ (and thus $R$) are independent of $\tau$. A random variable $\zeta_1$ is chosen from a uniformly random distribution between $0$ and $1$ ($\zeta_1\sim\mathcal{U}\left[0,1\right]$) and the time to the next event is calculated using
$\tau={R}^{-1}\log{\left(1/\zeta_1\right)}$.    
The rescaled  rates $a^{(n)\pm,r}_{\textrm{b(u)}}(i) = R^{-1}r_{\textrm{b(u)}}^{(n)\pm,i}/\tau$, $a^{(n)\pm,f}_{\textrm{b(u)}}(i) = R^{-1}f_{\textrm{b(u)}}^{(n)\pm,i}/\tau$ and $a^{(n)\pm,s}_{\textrm{b(u)}}(i) = R^{-1}s_{\textrm{b(u)}}^{(n)\pm,i}/\tau$ are concatenated in triplets for each force generator $n$, giving a list of potential states $a^j$ with $j\in [1,6N]$ which sum together to give ${\textstyle{\sum}}_{j=1}^{6N}a^{j}=1$ (Fig.~\ref{fig:stochastic_states}(b)). Choosing an independent random variable from a uniformly random distribution,  $\zeta_2\sim\mathcal{U}\left[0,1\right]$, the next state-changing event is determined as the first $j$ such that 
    ${\textstyle{\sum}}_{j'=1}^ja^{j'}>\zeta_2$.
Force generators in the upper ($n^+$) and lower ($n^-$) cortex have corresponding events $a^j$ where $j\in[1,3N]$ and $j\in[3N+1,6N]$ respectively. 
 
In order to calculate the spindle pole position, we implement a forward Euler approximation of (\ref{eq:polepos_nondim})  
which may be used to calculate the pole position at a time $t+\tau$,
\begin{equation}\label{eq:poleposstoch}
    z\left(t+\tau\right)=\left(1-\frac{\tau K}{\xi}\right)z\left(t\right)+\frac{\tau}{\xi}\left(\textstyle{\sum}_{n'=1}^{N'}y^{(n')+,i}_{\textrm{b}}\left(t\right)-\textstyle{\sum}_{n=1}^Ny^{(n)-,i}_{\textrm{b}}\left(t\right)\right).
\end{equation}
Here $n'$ and $N'$ are the equivalent of $n$ and $N$, introduced to separate the upper and lower cortex in this expression.

In summary, the state of the system at any instant can be described by a vector $\mathbf{X}(t)$ of size $\mathcal{N}=1+4N(M+1)$, comprising the spindle location plus the occupancies of $4N$ linkers ($N$ at each cortex, in bound or unbound states) in states of different lengths (over a scale discretized into $M+1$ elements) .  In principle such a system can be represented \citep{erban2020} by a chemical master equation for $p(\mathbf{x},t)$, the probability that $\mathbf{X}(t)=\mathbf{x}$, coupled (Langevin) stochastic differential equations for the elements of $\mathbf{X}$ and a chemical Fokker--Planck equation for $p(\mathbf{x},t)$.  The latter is a PDE of $\mathcal{N}+1$ dimensions.  This is distinct from the heavily reduced Fokker--Planck system (\ref{eq:FPE},\ref{eq:poleeom}) proposed by \cite{grill2005theory} that motivates the stochastic system illustrated in Fig.~\ref{fig:stochastic_states}.

\section{Reducing the Fokker--Planck equations to ODEs}
\label{A2}

To reduce the Fokker--Planck model to a system of ODEs, we rescale using the motor-protein-to-microtubule binding rate, writing ${t}=\tilde{t}/{\omega}_{\textrm{on}}$ and ${z}=\tilde{z}/{\omega}_{\textrm{on}}$. Then (\ref{eq:poleeom}) and (\ref{eq:FPE}) become
\begin{subequations}
\begin{align}
    {\xi}\tilde{z}_{\tilde{t}} = -\frac{K}{{\omega}_{\textrm{on}}}\tilde{z}-N\left(\int_0^{{y}_{\textrm{max}}} {y} {P}^{-}_{\textrm{b}}\textrm{d}{y}-\int_0^{{y}_{\textrm{max}}} {y} {P}^{+}_{\textrm{b}}\textrm{d}{y}\right), \label{eq:ODEeom} \\
    {\omega}_{\textrm{on}}{P}^{\pm}_{\textrm{b},\tilde{t}}+{J}^{\pm}_{\textrm{b},{y}} = {\omega}_{\textrm{on}}{P}^{\pm}_{\textrm{u}}-{\omega}_0e^{\gamma{y}}{P}^{\pm}_{\textrm{b}}, \label{eq:ODEPb} \\
    {\omega}_{\textrm{on}}{P}^{\pm}_{\textrm{u},\tilde{t}} + {J}^{\pm}_{\textrm{u},{y}} = -{\omega}_{\textrm{on}}{P}^{\pm}_{\textrm{u}}+{\omega}_0e^{\gamma{y}}{P}^{\pm}_{\textrm{b}}. \label{eq:ODEPu}
\end{align}
\end{subequations}
We develop an approximation to the oscillating spindle system for which ${\omega}_{\textrm{on}}\sim{\omega}_{0}\sim{D_{\mathrm{b}}}^{1/2}\sim{D_{\mathrm{u}}}^{1/2}\ll1$. To minimise the introduction of further notation, we expand our solutions in terms of the small order parameter ${\omega}_{\textrm{on}}$ and remain mindful that these parameters are taken to be of similar order. The range of extension values ${y}$ is split into three regions (Fig.~\ref{fig:PDEsol}(h)): region I over which ${P}_{\textrm{u}}^\pm$ is peaked around ${y}=0$ with a width ${D_{\mathrm{u}}}^{1/2}$; region III over which ${P}_{\textrm{b}}^\pm$ is peaked with a width of ${D_{\mathrm{b}}}^{1/2}$ but whose centre moves as ${y}_{\textrm{c}}=1\mp\tilde{z}_{\tilde{t}}$; and region II where advective terms dominate and the asymptotic limits of I and III are matched. Solutions for ${P}_{\textrm{u}}^\pm$ and ${P}_{\textrm{b}}^\pm$ will be determined in regions I and III respectively, followed by matching their asymptotic limits in region II to reveal the ODE system which governs the time evolution of the parameters.

\subsection{Region I}

In Region I, we seek solutions ${P}_{\textrm{u}}^\pm\sim {P}_{\textrm{u}0}^\pm+{\omega}_{\textrm{on}}{P}_{\textrm{u}1}^\pm+...$ where ${P}^\pm_{\textrm{u}0}$ is a quasi-static solution whose shape is static but whose amplitude varies slowly in time. 
We assume further that ${P}^\pm_{\textrm{b}}\sim{\omega}_{\textrm{on}}$ in this region (Fig.~\ref{fig:PDEsol}(h)). Here, it is observed that ${P}_{\textrm{u}}^\pm$ is sharply peaked about ${y}=0$ over a diffusive length-scale ${D_{\mathrm{u}}}^{1/2}$ (Fig.~\ref{fig:PDEsol}(h)). Thus, setting ${y}={D_{\mathrm{u}}}^{1/2}Y$ in (\ref{eq:ODEPu}) gives
\begin{equation}
    {\omega}_{\textrm{on}}{P}_{\textrm{u},\tilde{t}}^\pm-\Gamma\left(Y{P}_{\textrm{u}}^\pm+{P}_{\textrm{u},Y}^\pm\right)_{Y} = -{\omega}_{\textrm{on}}{P}_{\textrm{u}}^\pm+{\omega}_0e^{\left(\gamma{D_{\mathrm{u}}}^{1/2}Y\right)}{P}_{\textrm{b}}^\pm \label{eq:Pu_Y}
\end{equation}
with the boundary condition ${J}_{\textrm{u}}^\pm\left(\tilde{t},0\right)=0$. This boundary condition therefore becomes
\begin{equation}  \label{eq:Bounds}
{J}_{\textrm{u0}}^\pm\left(\tilde{t},0\right) +  {\omega}_{\textrm{on}}{J}_{\textrm{u1}}^\pm\left(\tilde{t},0\right) = 0
\end{equation}
where
\begin{subequations}  
\begin{align}
{J}_{\textrm{u0}}^\pm &= -{D_{\mathrm{u}}}^{1/2}\Gamma\left(Y{P}^\pm_{\textrm{u0}}+{P}^\pm_{\textrm{u0},Y}\right), \\
{J}_{\textrm{u1}}^\pm &= -{D_{\mathrm{u}}}^{1/2}\Gamma\left(Y{P}^\pm_{\textrm{u1}}+{P}^\pm_{\textrm{u1},Y}\right),
\end{align}
\end{subequations}
which are both individually zero at $Y=0$ due to (\ref{eq:Bounds}). To leading order in ${\omega}_{\textrm{on}}$, (\ref{eq:Pu_Y}) becomes
\begin{equation}            
\Gamma\left(Y{P}^\pm_{\textrm{u0}}+{P}^\pm_{\textrm{u0},Y}\right)_Y = 0
\end{equation}
which may be integrated to give
$\Gamma\left[Y{P}^\pm_{\textrm{u0}}+{P}^\pm_{\textrm{u0},Y}\right]^Y_0=0$.
Thus, due to boundary condition (\ref{eq:Bounds}),
\begin{equation}
 Y{P}^\pm_{\textrm{u0}}+{P}^\pm_{\textrm{u0},Y}=0,
\end{equation}
which gives
\begin{equation}\label{eq:Pu0RI}
{P}_{\textrm{u}0}^\pm=A^\pm(\tilde{t})e^{-\frac12Y^2} 
\end{equation}
as a solution, with $A^\pm\left(\tilde{t}\right)$ an amplitude which varies slowly in time. 

At $\mathcal{O}\left({\omega}_{\textrm{on}}\right)$, (\ref{eq:Pu_Y}) becomes
\begin{align}\label{eq:PuStep1}
    \Gamma\left(Y{P}_{\textrm{u}1}^\pm+{P}_{\textrm{u}1,Y}^\pm\right)_{Y}& = {P}_{\textrm{u}0,\tilde{t}}^\pm+{P}_{\textrm{u}0}^\pm. 
\end{align}
We highlight here the absence of the ${\omega}_0e^{\gamma{y}}{P}_{\textrm{b}}^\pm$ term as we have assumed that ${P}_{\textrm{b}}^\pm\sim{\omega}_{\textrm{on}}$ in this region. Then (\ref{eq:PuStep1}) may be integrated to
\begin{equation}
    \Gamma\left[Y{P}_{\textrm{u}1}^\pm+{P}_{\textrm{u}1,Y}^\pm\right]^Y_0 = \int_0^{Y}\left(A^\pm_{\tilde{t}}+A^\pm\right)e^{-\frac12Y^2}\textrm{d}Y 
\end{equation}
using the boundary conditions on flux (\ref{eq:Bounds}) at $Y=0$. As the $Y{P}_{\textrm{u1}}^\pm$ term will dominate the left-hand side as $Y\rightarrow\infty$, we may write
\begin{equation}
    \Gamma Y{P}_{\textrm{u}1}^\pm \approx \int_0^{\infty}\left(\left(A^\pm_{\tilde{t}}+A^\pm\right)e^{-\frac12Y^2}\right)\textrm{d}Y = \left(A^\pm_{\tilde{t}}+A^\pm\right)\sqrt{\frac\pi2}.
\end{equation}
Then,
\begin{align}
    {P}_{\textrm{u}1}^\pm &\approx \frac{1}{\Gamma Y}\left(A^\pm_{\tilde{t}}+A^\pm\right)\sqrt{\frac\pi2}, \quad \quad \left(Y\gg1\right).
\end{align}
Re-substitution of $Y={D_{\mathrm{u}}}^{-1/2}{y}$ gives
\begin{align}\label{eq:Pu1RI}
   {P}_{\textrm{u}1}^\pm  & \approx \frac{1}{\Gamma {y}}\left(A^\pm_{\tilde{t}}+A^\pm\right)\sqrt{\frac{\pi{D_{\mathrm{u}}}}{2}}
\end{align}
when ${D_{\mathrm{u}}}^{1/2} \ll {y}$.

Now that we have an expression for how the shape of the unbound force generator pdf varies in time in region I, we seek similar solutions for the bound generator pdf in region III.

\subsection{Region III}
In region III we seek solutions of the form ${P}_{\textrm{b}}^\pm\sim {P}_{\textrm{b}0}^\pm+{\omega}_{\textrm{on}}{P}_{\textrm{b}1}^\pm+...$. Here, the pdf ${P}_{\textrm{b}}^\pm$ is sharply peaked about ${y}_{\textrm{c}}=1\mp\tilde{z}_{\tilde{t}}$ over a diffusive length-scale ${D_{\mathrm{b}}}^\frac{1}{2}$ (Fig.~\ref{fig:PDEsol}(h)). Both ${D_{\mathrm{b}}}^{1/2}$ and ${\omega}_{\textrm{on}}$ are assumed to be small parameters of similar order. Thus, in this region about the peak of ${P}_{\textrm{b}}^\pm$, we set ${y}=1\mp\tilde{z}_{\tilde{t}}+{D_{\mathrm{b}}}^\frac{1}{2}\hat{Y}$. Noting that
\begin{equation}
\frac{\partial}{\partial \tilde{t}}\rightarrow\pm{D_{\mathrm{b}}}^{-1/2}\tilde{z}_{\tilde{t}\tilde{t}}\frac{\partial}{\partial \hat{Y}}+\frac{\partial}{\partial \tilde{t}}, \quad
    \frac{\partial}{\partial{y}}\rightarrow{D_{\mathrm{b}}}^{-1/2}\frac{\partial}{\partial\hat{Y}}
\end{equation}
(\ref{eq:ODEPb}) becomes, to leading order,
\begin{align} \label{eq:PbODEY}
    {\omega}_{\textrm{on}}\left({P}^\pm_{\textrm{b},\tilde{t}}\pm{D_{\mathrm{b}}}^{-1/2}\tilde{z}_{\tilde{t}\tilde{t}}{P}^\pm_{\textrm{b},\hat{Y}}\right)-\left({D_{\mathrm{b}}}^{-1/2}{v}^\pm_{\textrm{b}}{P}_{\textrm{b}}^\pm+{P}^\pm_{\textrm{b},\hat{Y}}\right)_{\hat{Y}}={\omega}_{\textrm{on}}{P}_{\textrm{u}}^\pm-{\omega}_{0}e^{\gamma\left(1\mp\tilde{z}_{\tilde{t}}\right)}{P}_{\textrm{b}}^\pm.
\end{align}
Recalling 
(\ref{eq:FPEvel}),
(\ref{eq:PbODEY}) may be further simplified to
\begin{align} \label{eq:PbODEY2}
    {\omega}_{\textrm{on}}\left({P}^\pm_{\textrm{b},\tilde{t}}\pm{D_{\mathrm{b}}}^{-1/2}\tilde{z}_{\tilde{t}\tilde{t}}{P}^\pm_{\textrm{b},\hat{Y}}\right)-\left(\hat{Y}{P}_{\textrm{b}}^\pm+{P}^\pm_{\textrm{b},\hat{Y}}\right)_{\hat{Y}}={\omega}_{\textrm{on}}{P}_{\textrm{u}}^\pm-{\omega}_{0}e^{\gamma\left(1\mp\tilde{z}_{\tilde{t}}\right)}{P}_{\textrm{b}}^\pm.
\end{align}
Substituting the expansion ${P}_{\textrm{b}}^\pm\approx{P}_{\textrm{b0}}^\pm+{\omega}_{\textrm{on}}{P}_{\textrm{b1}}^\pm+\hdots$, to first order (\ref{eq:PbODEY2}) becomes
\begin{align}
    \left(\left(\hat{Y}\mp{\omega}_{\textrm{on}}{D_{\mathrm{b}}}^{-1/2}\tilde{z}_{\tilde{t}\tilde{t}}\right){P}^\pm_{\textrm{b}0}+{P}^\pm_{\textrm{b}0,\hat{Y}}\right)_{{Y}}=0
\end{align}
taking ${{\omega}_{\textrm{on}}}/{{D_{\mathrm{b}}}^{1/2}}$ to be $\mathcal{O}\left(1\right)$. By integration,
\begin{align}\label{eq:Pbintegrate}
    \left({Y}\mp{\omega}_{\textrm{on}}{D_{\mathrm{b}}}^{-1/2}\tilde{z}_{\tilde{t}\tilde{t}}\right){P}^\pm_{\textrm{b}0}-{P}^\pm_{\textrm{b}0,{Y}}=C\left(\tilde{t}\right)
\end{align}
for $C\left(\tilde{t}\right)$ some constant of integration. The boundary condition ${J}^\pm_{\textrm{b}}\left(\tilde{t},{y}={y}_{\textrm{max}}\right)=0$ becomes ${J}^\pm_{\textrm{b}}\left(\tilde{t},\hat{Y}\rightarrow\infty\right)\rightarrow0$, while ${J}^\pm_{\textrm{b}}\left(\tilde{t},{y}=0\right)=0$ becomes ${J}^\pm_{\textrm{b}}\left(\tilde{t},\hat{Y}\rightarrow-\infty\right)\rightarrow0$. Then
\begin{equation}\label{eq:RIIIBounds}
{J}^\pm_{\textrm{b0}}\left(\tilde{t},\hat{Y}\rightarrow-\infty\right)+{\omega}_{\textrm{on}}{J}^\pm_{\textrm{b1}}\left(\tilde{t},\hat{Y}\rightarrow-\infty\right)\rightarrow0
\end{equation}
where
\begin{subequations}
\begin{align}
{J}^\pm_{\textrm{b0}}={D_{\mathrm{b}}}^{1/2}\left(\hat{Y}{P}^\pm_{\textrm{b0}}-{P}^\pm_{\textrm{b0},\hat{Y}}\right), \\
{J}^\pm_{\textrm{b1}}={D_{\mathrm{b}}}^{1/2}\left(\hat{Y}{P}^\pm_{\textrm{b1}}-{P}^\pm_{\textrm{b1},\hat{Y}}\right)
\end{align}
\end{subequations}
which must both separately also tend to zero as $\hat{Y}\rightarrow-\infty$. Using this, (\ref{eq:Pbintegrate}) may be rewritten as
\begin{equation}
    {D_{\mathrm{b}}}^{-1/2}{J}^\pm_{\textrm{b0}}\mp{\omega}_{\textrm{on}}{D_{\mathrm{b}}}^{-1/2}\tilde{z}_{\tilde{t}\tilde{t}}{P}^\pm_{\textrm{b0}}=C\left(\tilde{t}\right).
\end{equation}
Making the assumption that ${P}^\pm_{\textrm{b}0}\rightarrow0$ as $\hat{Y}\rightarrow-\infty$, which enforces (\ref{eq:RIIIBounds}) at leading order, then $C\left(\tilde{t}\right)=0$ and
\begin{align}
    {P}^\pm_{\textrm{b}0,\hat{Y}}=-\left(\hat{Y}\mp{\omega}_{\textrm{on}}{D_{\mathrm{b}}}^{-1/2}\tilde{z}_{\tilde{t}\tilde{t}}\right){P}^\pm_{\textrm{b}0}. \label{eq:RIIIstep}
\end{align}
Integrating (\ref{eq:RIIIstep}) gives the solution
\begin{align}\label{eq:Pb0_sol}
{P}^\pm_{\textrm{b}0}=\tilde{B}^\pm\left(\tilde{t}\right)e^{\left(-\frac12\hat{Y}^2\pm{\omega}_{\textrm{on}}{D_{\mathrm{b}}}^{-1/2}\tilde{z}_{\tilde{t}\tilde{t}}\hat{Y}\right)}=B^\pm\left(\tilde{t}\right)e^{\left(-\frac12\left(\hat{Y}\mp{\omega}_{\textrm{on}}{D_{\mathrm{b}}}^{-1/2}\tilde{z}_{\tilde{t}\tilde{t}}\right)^2\right)}
\end{align}
for some $\tilde{B}^\pm\left(\tilde{t}\right)$ and $B^{\pm}\left(\tilde{t}\right)$, where $B^\pm\left(\tilde{t}\right)$ describes the amplitude of the peak of the pdf (subject to smaller corrections) which varies in time. Returning to (\ref{eq:PbODEY2}), with ${P}^\pm_{\textrm{u}}\ll{P}^\pm_{\textrm{b}}$, then to $\mathcal{O}\left({\omega}_{\textrm{on}}\right)$
\begin{align}
    \left(\left(\hat{Y}\mp{\omega}_{\textrm{on}}{D_{\mathrm{b}}}^{-1/2}\tilde{z}_{\tilde{t}\tilde{t}}\right){P}^\pm_{\textrm{b}1}+{P}^\pm_{\textrm{b}1,\hat{Y}}\right)_{\hat{Y}}={P}^\pm_{\textrm{b}0,\tilde{t}}+\frac{{\omega}_{0}e^{\gamma\left(1\mp\tilde{z}_{\tilde{t}}\right)}}{{\omega}_{\textrm{on}}}{P}^\pm_{\textrm{b}0},
\end{align}
which may be rewritten as
\begin{align}\label{eq:Pb1_step1}
    \left({D_{\mathrm{b}}}^{-1/2}{J}^\pm_{\textrm{b1}}\mp{\omega}_{\textrm{on}}{D_{\mathrm{b}}}^{-1/2}\tilde{z}_{\tilde{t}\tilde{t}}{P}^\pm_{\textrm{b1}}\right)_{\hat{Y}}=
    {P}^\pm_{\textrm{b}0,\tilde{t}}+\frac{{\omega}_{0}e^{\gamma\left(1\mp\tilde{z}_{\tilde{t}}\right)}}{{\omega}_{\textrm{on}}}{P}^\pm_{\textrm{b}0}.
\end{align}
Thus using (\ref{eq:Pb0_sol}) in (\ref{eq:Pb1_step1}) and integrating gives
\begin{multline*}
    \left[{D_{\mathrm{b}}}^{-1/2}{J}^\pm_{\textrm{b1}}\mp {\omega}_{\textrm{on}}{D_{\mathrm{b}}}^{-1/2}\tilde{z}_{\tilde{t}\tilde{t}}{P}^\pm_{\textrm{b1}}\right]^\infty_{{Y}}=\\
    \int^\infty_{{Y}}\left(B^\pm_{\tilde{t}}\pm{\omega}_{\textrm{on}}{D_{\mathrm{b}}}^{-1/2}\tilde{z}_{\tilde{t}\tilde{t}\tilde{t}}\left({Y}\mp{\omega}_{\textrm{on}}{D_{\mathrm{b}}}^{-1/2}\tilde{z}_{\tilde{t}\tilde{t}} \right)B^\pm\right)e^{\left(-\frac12\left({Y}\mp{\omega}_{\textrm{on}}{D_{\mathrm{b}}}^{-1/2}\tilde{z}_{\tilde{t}\tilde{t}}\right)^2\right)} \nonumber \\
 +\left(\frac{{\omega}_{0}e^{\gamma\left(1\mp\tilde{z}_{\tilde{t}}\right)}}{{\omega}_{\textrm{on}}}B^\pm\right)e^{\left(-\frac12\left({Y}\mp{\omega}_{\textrm{on}}{D_{\mathrm{b}}}^{-1/2}\tilde{z}_{\tilde{t}\tilde{t}}\right)^2\right)}\textrm{d}{Y}.
\end{multline*}
By assuming that ${P}^\pm_{\textrm{b}1}\rightarrow0$ as $\hat{Y}\rightarrow\infty$, which enforces boundary condition ${J}^\pm_{\textrm{b1}}\rightarrow0$ as $\hat{Y}\rightarrow\infty$, then
\begin{multline*}
    -{D_{\mathrm{b}}}^{-1/2}{J}^\pm_{\textrm{b1}}\pm{\omega}_{\textrm{on}}{D_{\mathrm{b}}}^{-1/2}\tilde{z}_{\tilde{t}\tilde{t}}{P}^\pm_{\textrm{b1}}= \\ \int^\infty_{\hat{Y}} \left(B^\pm_{\tilde{t}}\pm{\omega}_{\textrm{on}}{D_{\mathrm{b}}}^{-1/2}\tilde{z}_{\tilde{t}\tilde{t}\tilde{t}}\left(\hat{Y}\mp{\omega}_{\textrm{on}}{D_{\mathrm{b}}}^{-1/2}\tilde{z}_{\tilde{t}\tilde{t}} \right)B^\pm\right)e^{\left(-\frac12\left(\hat{Y}\mp{\omega}_{\textrm{on}}{D_{\mathrm{b}}}^{-1/2}\tilde{z}_{\tilde{t}\tilde{t}}\right)^2\right)} \nonumber \\
 +\left(\frac{{\omega}_{0}e^{\gamma\left(1\mp\tilde{z}_{\tilde{t}}\right)}}{{\omega}_{\textrm{on}}}B^\pm\right)e^{\left(-\frac12\left(\hat{Y}\mp{\omega}_{\textrm{on}}{D_{\mathrm{b}}}^{-1/2}\tilde{z}_{\tilde{t}\tilde{t}}\right)^2\right)}\textrm{d}\hat{Y}. \label{eq:RIIIIntegs}
\end{multline*}
The left-hand side (LHS) may be rewritten 
\begin{equation}\label{eq:LHS}
    \textrm{LHS} = -\left(\hat{Y}\mp{D_{\mathrm{b}}}^{-1/2}{\omega}_{\textrm{on}}\tilde{z}_{\tilde{t}\tilde{t}}\right){P}^\pm_{\textrm{b1}}+{P}^\pm_{\textrm{b1},\hat{Y}}
\end{equation}
and so in the limit $\hat{Y}\rightarrow-\infty$, (\ref{eq:LHS}) is dominated by the $\hat{Y}{P}^\pm_{\textrm{b}1}$ term. Rearranging the right-hand side gives, in this limit,
\begin{multline*}
   -\hat{Y}{P}^\pm_{\textrm{b}1}\sim\Biggl(B^\pm_{\tilde{t}}+\frac{{\omega}_{0}e^{\gamma\left(1\mp\tilde{z}_{\tilde{t}}\right)}}{{\omega}_{\textrm{on}}} B^\pm\Biggr)\int^\infty_{-\infty}e^{\left(-\frac12\left(\hat{Y}\mp{\omega}_{\textrm{on}}{D_{\mathrm{b}}}^{-1/2}\tilde{z}_{\tilde{t}\tilde{t}}\right)^2\right)}\textrm{d}\hat{Y} \nonumber \\
     \pm{\omega}_{\textrm{on}}{D_{\mathrm{b}}}^{-1/2}\tilde{z}_{\tilde{t}\tilde{t}\tilde{t}}B^\pm\int^{\infty}_{-\infty}\left(\hat{Y}\mp{\omega}_{\textrm{on}}{D_{\mathrm{b}}}^{-1/2}\tilde{z}_{\tilde{t}\tilde{t}}\right)e^{\left(-\frac12\left(\hat{Y}\mp{\omega}_{\textrm{on}}{D_{\mathrm{b}}}^{-1/2}\tilde{z}_{\tilde{t}\tilde{t}}\right)^2\right)}\textrm{d}\hat{Y}. 
\end{multline*}
The second integral vanishes, while the first integral can be evaluated and thus, as $\hat{Y}\rightarrow-\infty$,
\begin{align}\label{eq:Pb1RIII}
     {P}^\pm_{\textrm{b}1}\sim-\frac{\sqrt{2\pi}}{\hat{Y}}\left(B^\pm_{\tilde{t}}+\frac{{\omega}_{0}e^{\gamma\left(1\mp\tilde{z}_{\tilde{t}}\right)}}{{\omega}_{\textrm{on}}}B^\pm\right)=\frac{\sqrt{2\pi{D_{\mathrm{b}}}}}{1-{y}\mp\tilde{z}_{\tilde{t}}}\left(B^\pm_{\tilde{t}}+\frac{{\omega}_{0}e^{\gamma\left(1\mp\tilde{z}_{\tilde{t}}\right)}}{{\omega}_{\textrm{on}}}B^\pm\right).
\end{align}

The asymptotic limits (\ref{eq:Pu1RI}) and (\ref{eq:Pb1RIII}) as $Y\rightarrow\infty$ and $\hat{Y}\rightarrow-\infty$ respectively will now be matched inside region II.

\subsection{Region II}
In region II, advection terms dominate. These `sweep' the bound force generators toward the peak of ${P}^\pm_{\textrm{b}}$ such that bound force generators will tend to have elastic linkers with an extension ${y}_{\textrm{c}}=1\mp\tilde{z}_{\tilde{t}}$, and the unbound force generators toward the peak of ${P}^\pm_{\textrm{u}}$ such that unbound force generators will tend to have an elastic linker with zero extension. Given that the pdfs are peaked in regions I and II, ${P}^\pm_{\textrm{b},{y}}$ and ${P}^\pm_{\textrm{u},{y}}$ are both relatively small in region II, being given by the small correction terms ${\omega}_{\textrm{on}}{P}^\pm_{\textrm{u1}}$ and ${\omega}_{\textrm{on}}{P}^\pm_{\textrm{b1}}$, expressions for which we have determined in the limits $Y\rightarrow\infty$ and $\hat{Y}\rightarrow-\infty$ respectively. Then together, using (\ref{eq:FPEb}),
 $   {J}_{\textrm{b}}^\pm={v}^\pm_{\textrm{b}}{P}_{\textrm{b}}^\pm-{D_{\mathrm{b}}}{P}^\pm_{\textrm{b},{y}} \approx {v}^\pm_{\textrm{b}}{P}_{\textrm{b}}^\pm $,
and so substitution of ${P}^\pm_{\textrm{b}}\approx{\omega}_{\textrm{on}}{P}^\pm_{\textrm{b1}}$ when  $\hat{Y}\rightarrow-\infty$ and (\ref{eq:FPEvel}) returns
\begin{align}\label{eq:JbRII}
    {J}_{\textrm{b}}^\pm\approx v^\pm_{\textrm{b}}P_{\textrm{b}}^\pm\approx\sqrt{2\pi{D_{\mathrm{b}}}}\left({\omega}_{0}e^{\gamma\left(1\mp\tilde{z}_{\tilde{t}}\right)}B^\pm+{\omega}_{\textrm{on}}B^\pm_{\tilde{t}}\right).
    \end{align}
Continuing, using (\ref{eq:FPEu}),
 $   {J}_{\textrm{u}}^\pm= -\Gamma\left({y}{P}_{\textrm{u}}^\pm+{D_{\mathrm{u}}}{P}^\pm_{\textrm{u},{y}}\right) \approx -\Gamma{y}{P}^\pm_{\textrm{u}}$,
and so substitution of ${P}^\pm_{\textrm{u}}\approx{\omega}_{\textrm{on}}{P}^\pm_{\textrm{u1}}$ when $Y\rightarrow\infty$ returns
\begin{align}\label{eq:JuRII}
    {J}_{\textrm{u}}^\pm\approx-\Gamma{y}{P}^\pm_{\textrm{u}}\approx-{\omega}_{\textrm{on}}\sqrt{\frac{\pi{D_{\mathrm{u}}}}{2}}\left(A^\pm+A^\pm_{\tilde{t}}\right).
\end{align}
By the form of (\ref{eq:JbRII}) and (\ref{eq:JuRII}), ${J}^\pm_{\textrm{b},{y}}=0$ and ${J}^\pm_{\textrm{u},{y}}=0$ to leading order, and therefore (\ref{eq:JbRII}) and (\ref{eq:JuRII}) are valid across the whole of region II. It follows that
\begin{equation}
    {J}_{\textrm{b}}^\pm+{J}_{\textrm{u}}^\pm=\sqrt{D_{\textrm{u}}}\omega_{\textrm{on}}\left[\sqrt{\frac{2\pi D_{\textrm{b}}}{D_{\textrm{u}}}}\left(\frac{\omega_0}{\omega_{\textrm{on}}}e^{\gamma\left(1\mp\tilde{z}_{\tilde{t}}\right)}B^\pm+B^\pm_{\tilde{t}}\right)-\sqrt{\frac\pi2}\left(A^\pm+A^\pm_{\tilde{t}}\right)\right]
\end{equation}
is a constant across region II.  As demonstrated in Online Resource 2(a,b), detailed balance ($J_{\mathrm{b}}^\pm+J_{\mathrm{u}}^\pm=0$) does not hold in this region.

\subsection{Matching solutions: Regions I-II}
In region I it was assumed that $P_{\textrm{b}}^\pm$ was sufficiently small ($\mathcal{O}\left(\omega_{\textrm{on}}\right)$) that its dynamics could be neglected to leading order. By (\ref{eq:JbRII}) it can be estimated that $P^\pm_{\textrm{b}}$ is of magnitude $\omega_{\textrm{on}}\sqrt{D_{\textrm{u}}}$, where it has been assumed that $D_{\textrm{b}}\sim D_{\textrm{u}}$. Then letting \hbox{$P^\pm_{\textrm{b}}=\omega_{\textrm{on}}\sqrt{D_{\textrm{u}}}\hat{P}_{\textrm{b}}^\pm$} in (\ref{eq:ODEPu}) when $y=\sqrt{D_{\textrm{u}}}Y$ results in expressions for $P_{\textrm{u0}}^\pm$ and $P^\pm_{\textrm{u1}}$ which are unchanged from (\ref{eq:Pu0RI}) and (\ref{eq:Pu1RI}) respectively. However, (\ref{eq:ODEPb}) becomes
\begin{equation}\label{eq:PbRI-II}
\omega_{\textrm{on}}\hat{P}_{\textrm{b},\tilde{t}}^\pm+\frac{1}{\sqrt{D_{\textrm{u}}}}\left(\left(1\mp\tilde{z}_{\tilde{t}}-\sqrt{D_{\textrm{u}}}Y\right)\hat{P}^\pm_{\textrm{b}}-\sqrt{D_{\textrm{b}}}\hat{P}^\pm_{\textrm{b},Y}\right)_{Y}=\frac{1}{\sqrt{D_{\textrm{u}}}}P^\pm_{\textrm{u}}-\omega_0 e^{\gamma\sqrt{D_{\textrm{u}}}Y}\hat{P}_{\textrm{b}}^\pm.
\end{equation}
To leading order, (\ref{eq:PbRI-II}) becomes
\begin{equation}
    \left(1\mp\tilde{z}_{\tilde{t}}\right)\hat{P}^\pm_{\textrm{b},Y}=P_{\textrm{u0}}^\pm=A^\pm e^{-Y^2/2}.
\end{equation}
Then, integrating over $Y$ from $Y=0$ to $Y\rightarrow\infty$, 
\begin{equation}
    \left(1\mp\tilde{z}_{\tilde{t}}\right)\hat{P}^\pm_{\textrm{b}}=\sqrt{\frac\pi2}A^\pm.
\end{equation}
Assuming that $v_{\textrm{b}}^\pm\approx1\mp\tilde{z}_{\tilde{t}}$ to leading order, then
\begin{equation}\label{eq:JbRI-II}
    J_{\textrm{b}}^\pm\approx\omega_{\textrm{on}}\sqrt{\frac{D_{\textrm{u}}\pi}{2}}A^\pm.
\end{equation}
Since $J_{\textrm{b}}^\pm$ is independent of $y$ in region II (Online Resource 2a,b), then we match (\ref{eq:JbRII}) to (\ref{eq:JbRI-II}), resulting in
\begin{equation}\label{eq:B-A-RI-RII}
    \frac{\omega_0}{\omega_{\textrm{on}}}e^{\gamma\left(1\mp\tilde{z}_{\tilde{t}}\right)}B^\pm+B^\pm_{\tilde{t}}=\frac12\sqrt{\frac{D_{\textrm{u}}}{D_{\textrm{b}}}}A^\pm.
\end{equation}
We now perform the same analysis on the boundary of regions II and III.

\subsection{Matching solutions: Regions II-III}
In region III it was assumed that $P_{\textrm{u}}^\pm$ was sufficiently small that its dynamics could be neglected to leading order. By (\ref{eq:JuRII}) it can be estimated that $P^\pm_{\textrm{u}}$ is of magnitude $\omega_{\textrm{on}}\sqrt{D_{\textrm{u}}}$. Then letting \hbox{$P^\pm_{\textrm{u}}=\omega_{\textrm{on}}\sqrt{D_{\textrm{u}}}\hat{P}_{\textrm{u}}^\pm$} in (\ref{eq:ODEPb}) when $y=1\mp\tilde{z}_{\tilde{t}}+\sqrt{D_{\textrm{b}}}\hat{Y}$ results in expressions for $P_{\textrm{b0}}^\pm$ and $P^\pm_{\textrm{b1}}$ which are unchanged from (\ref{eq:Pb0_sol}) and (\ref{eq:Pb1RIII}) respectively. However, (\ref{eq:ODEPu}) becomes
\begin{align}
\omega_{\textrm{on}}\left(\hat{P}_{\textrm{u},\tilde{t}}^\pm\pm\frac{\tilde{z}_{\tilde{t}\tilde{t}}}{\sqrt{D_{\textrm{b}}}}\hat{P}^\pm_{\textrm{u},\hat{Y}}\right)-\Gamma&\left(\frac{1}{\sqrt{D_{\textrm{b}}}}\left(1\mp\tilde{z}_{\tilde{t}}+\sqrt{D_{\textrm{b}}}\hat{Y}\right)\hat{P}^\pm_{\textrm{u}}+\frac{D_{\textrm{u}}}{D_{\textrm{b}}}\hat{P}^\pm_{\textrm{u},\hat{Y}}\right)_{\hat{Y}} \nonumber \\
&=-\omega_{\textrm{on}}\hat{P}^\pm_{\textrm{u}}+\frac{\omega_0}{\omega_{\textrm{on}}\sqrt{D_{\textrm{u}}}}e^{\gamma\left(1\mp\tilde{z}_{\tilde{t}}+\sqrt{D_{\textrm{b}}}\hat{Y}\right)}P^\pm_{\textrm{b}}.\label{eq:PuRII-III}
\end{align}
To leading order,
\begin{equation}
    -\frac{\Gamma}{\sqrt{D_{\textrm{b}}}}\left(1\mp\tilde{z}_{\tilde{t}}\right)\hat{P}_{\textrm{u},\hat{Y}}^\pm=\frac{\omega_0}{\omega_{\textrm{on}}\sqrt{D_{\textrm{u}}}}e^{\gamma\left(1\mp\tilde{z}_{\tilde{t}}\right)}P_{\textrm{b0}}^\pm=\frac{\omega_0}{\omega_{\textrm{on}}\sqrt{D_{\textrm{u}}}}e^{\gamma\left(1\mp\tilde{z}_{\tilde{t}}\right)}B^\pm e^{-\frac12\left(\hat{Y}\mp\omega_{\textrm{on}}D_{\textrm{b}}^{-1/2}\tilde{z}_{\tilde{t}\tilde{t}}\right)^2}
\end{equation}
where we have used that $e^{\gamma\left(1\mp\tilde{z}_{\tilde{t}}+\sqrt{D_{\textrm{b}}}\hat{Y}\right)}=e^{\gamma\left(1\mp\tilde{z}_{\tilde{t}}\right)}e^{\gamma\sqrt{D_{\textrm{b}}}\hat{Y}}\approx e^{\gamma\left(1\mp\tilde{z}_{\tilde{t}}\right)}$ as $\sqrt{D_{\textrm{b}}}$ is a small parameter. Integrating from $\hat{Y}\rightarrow-\infty$ to $\hat{Y}\rightarrow\infty$ gives
\begin{equation}
        \Gamma\left(1\mp\tilde{z}_{\tilde{t}}\right)\hat{P}_{\textrm{u}}^\pm=\sqrt{\frac{D_{\textrm{b}}}{D_{\textrm{u}}}}\frac{\omega_0}{\omega_{\textrm{on}}}e^{\gamma\left(1\mp\tilde{z}_{\tilde{t}}\right)}B^\pm \sqrt{2\pi}.
\end{equation}
 Taking $y\approx1\mp\tilde{z}_{\tilde{t}}$ to leading order, then
\begin{equation}
    J^\pm_{\textrm{u}}\approx-\sqrt{2\pi D_{\textrm{b}}}\omega_0e^{\gamma\left(1\mp\tilde{z}_{\tilde{t}}\right)}B^\pm. \label{eq:JuRII-RIII}
\end{equation}
Since $J_{\textrm{u}}^\pm$ is independent of $y$ in region II (Online Resource 2a,b), then we match (\ref{eq:JuRII}) to (\ref{eq:JuRII-RIII}) resulting in
\begin{equation}\label{eq:B-A-RII-RIII}
    \frac{\omega_0}{\omega_{\textrm{on}}}e^{\gamma\left(1\mp\tilde{z}_{\tilde{t}}\right)}B^\pm=\frac12\sqrt{\frac{D_{\textrm{u}}}{D_{\textrm{b}}}}\left(A^\pm+A^\pm_{\tilde{t}}\right).
\end{equation}

\subsection{Combining the whole system}

We may now use expressions (\ref{eq:Pu0RI}), (\ref{eq:Pu1RI}) for ${P}_{\textrm{b}}^\pm$ and (\ref{eq:Pb0_sol}), (\ref{eq:Pb1RIII}) for ${P}_{\textrm{u}}^\pm$, and their coupling in region II (\ref{eq:B-A-RI-RII}, \ref{eq:B-A-RII-RIII}) to close the system. Recalling (\ref{eq:Pu+Pb=1}),
then to leading order 
\begin{equation}
\int_0^\infty\left(A^\pm e^{-\frac{1}{2{D_{\mathrm{u}}}}{y}^2}+B^\pm e^{\left(-\frac{1}{2{D_{\mathrm{b}}}}\left({y}-1\pm\tilde{z}_{\tilde{t}}\mp{\omega}_{\textrm{on}}\tilde{z}_{\tilde{t}\tilde{t}}\right)^2\right)}\right)\textrm{d}{y}=1.
\end{equation}
The first term of this integral is easily evaluated, while the second term is more complex. Consider only the leading-order terms of the exponent, due to ${\omega}_{\textrm{on}}$ being a small order parameter. Then
\begin{equation}\label{eq:GaussCurve}
\int_0^\infty B^\pm e^{\left(-\frac{1}{2{D_{\mathrm{b}}}}\left({y}-1\pm\tilde{z}_{\tilde{t}}\mp{\omega}_{\textrm{on}}\tilde{z}_{\tilde{t}\tilde{t}}\right)^2\right)}\textrm{d}{y}\approx\int_0^\infty B^\pm e^{\left(-\frac{1}{2{D_{\mathrm{b}}}}\left({y}-{y}_{\textrm{c}}\right)^2\right)}\textrm{d}{y}
\end{equation}
which we know is a peak contained within region III. That is, we integrate over the Gaussian, which does not intersect ${y}=0$ with any value of significance at leading order. Using this logic, the integral (\ref{eq:GaussCurve}) may be evaluated and thus
\begin{equation}\label{eq:balance}
A^\pm\sqrt{\frac{\pi{D_{\mathrm{u}}}}{2}}+B^\pm\sqrt{2\pi{D_{\mathrm{b}}}}=1.
\end{equation} 
This can be used to eliminate $A^\pm$ from (\ref{eq:B-A-RI-RII}) to give
\begin{align}\label{eq:Bt}
    \sqrt{2\pi{D_{\mathrm{b}}}}B^\pm_{\tilde{t}} =1 -\sqrt{2\pi{D_{\mathrm{b}}}}\left(\frac{{\omega}_{0}}{\omega_{\textrm{on}}}e^{\gamma\left(1\mp\tilde{z}_{\tilde{t}}\right)}+1\right)B^\pm.
\end{align}
(\ref{eq:balance}) may similarly be used in (\ref{eq:B-A-RII-RIII}) to return (\ref{eq:Bt}).

Equation~(\ref{eq:Bt}) predicts that $\sqrt{2\pi{D_{\mathrm{b}}}}B^\pm$ relaxes to  ${{\omega}_{\textrm{on}}} / \{{\omega}_{\textrm{on}}+{\omega}_{0}e^{\gamma\left(1\mp\tilde{z}_{\tilde{t}}\right)}\}$ and (\ref{eq:balance}) predicts that $\sqrt{{\pi{D_{\mathrm{u}}}}/{2}}A^\pm$ relaxes to ${{\omega}_{0}e^{\gamma\left(1\mp\tilde{z}_{\tilde{t}}\right)}}/\{{\omega}_{\textrm{on}}+{\omega}_{0}e^{\gamma\left(1\mp\tilde{z}_{\tilde{t}}\right)}\}$ provided $\tilde{z}$ does not change too rapidly.

Further to this, ${P}_{\textrm{b}}^\pm={P}_{\textrm{b0}}^\pm+{\omega}_{\textrm{on}}{P}_{\textrm{b1}}^\pm+\hdots$ can be put into (\ref{eq:ODEeom}) to obtain a leading-order equation for the motion of the spindle pole. This requires the evaluation of
\begin{equation}
\int_0^{{y}_{\textrm{max}}}{y}{P}^\pm_{\textrm{b}}\textrm{d}{y} \sim \int_0^\infty{y}B^\pm e^{\left(-\frac1{2{D_{\mathrm{b}}}}\left({y}-{y}_{\textrm{c}}\right)\right)^2}\textrm{d}{y}+...
\end{equation}
where we assume ${y}_{\textrm{max}}$ is sufficiently large that it exceeds the bounds of region III and can thus be taken as ${y}_{\textrm{max}}\rightarrow\infty$. Again we let ${y}={y}_{\textrm{c}}+{D_{\mathrm{b}}}^{1/2}\hat{Y}$, which is where ${P}^\pm_{\textrm{b0}}$ has a significant value. Then
\begin{equation}
\int_0^\infty{y}B^\pm e^{\left(-\frac1{2{D_{\mathrm{b}}}}\left({y}-{y}_{\textrm{c}}\right)^2\right)}\textrm{d}{y}\sim\int_{-\infty}^{\infty}\left({y}_{\textrm{c}}+{D_{\mathrm{b}}}^{1/2}\hat{Y}\right)B^\pm e^{-\frac12\hat{Y}^2}{D_{\mathrm{b}}}^{1/2}\textrm{d}\hat{Y},
\end{equation}
which to leading order becomes
\begin{align}
\int_{-\infty}^{\infty}\left({y}_{\textrm{c}}+{D_{\mathrm{b}}}^{1/2}\hat{Y}\right)B^\pm e^{-\frac12\hat{Y}^2}{D_{\mathrm{b}}}^{1/2}\textrm{d}\hat{Y}&\sim{D_{\mathrm{b}}}^{1/2}{y}_{\textrm{c}}B^\pm\int_{-\infty}^{\infty}e^{-\frac12\hat{Y}^2}\textrm{d}\hat{Y}\nonumber \\
& ={y}_{\textrm{c}}B^{\pm}\sqrt{2\pi{D_{\mathrm{b}}}}+... .
\end{align}
Recalling that ${y}_{\textrm{c}}=1\mp\tilde{z}_{\tilde{t}}$, (\ref{eq:ODEeom}) becomes
\begin{align}
{\xi}\tilde{z}_{\tilde{t}} = -\frac{K}{{\omega}_{\textrm{on}}}\tilde{z}-N\sqrt{2\pi{D_{\mathrm{b}}}}\left(\left(1+\tilde{z}_{\tilde{t}}\right)B^--\left(1-\tilde{z}_{\tilde{t}}\right)B^+\right)
\end{align}
and thus
\begin{align}
{\xi}\tilde{z}_{\tilde{t}} = -\frac{K}{{\omega}_{\textrm{on}}}\tilde{z}-N\sqrt{2\pi{D_{\mathrm{b}}}}\left(B^-+B^+\right)\tilde{z}_{\tilde{t}}-N\sqrt{2\pi{D_{\mathrm{b}}}}\left(B^--B^+\right). 
\end{align}
This can be rewritten as
\begin{align}
    \left(\hat{\xi}+\hat{B}^++\hat{B}^-\right)\tilde{z}_{\tilde{t}}+\hat{K}\tilde{z}=\hat{B}^+-\hat{B}^-, \label{eq:ODE1_A}
\end{align}
where $\hat{B}^\pm=\sqrt{2\pi{D_{\mathrm{b}}}}B^\pm$, $\hat{K}={K}/\{{\omega}_{\textrm{on}}N\}$, and $\hat{\xi}={{\xi}}/{N}$. 
Recalling (\ref{eq:Bt}) which may be alternatively written as
\begin{align}
    \left(1+\rho e^{\gamma\left(1\mp\tilde{z}_{\tilde{t}}\right)}\right)\hat{B}^\pm+\hat{B}^\pm_{\tilde{t}} = 1 \label{eq:ODE2_A}
\end{align}
where $\rho={\omega}_0/{\omega}_{\textrm{on}}$, the coupled system (\ref{eq:FPEb}), (\ref{eq:FPEu}), and (\ref{eq:poleeom}) is reduced to solving (\ref{eq:ODE1_A}, \ref{eq:ODE2_A}) along with initial conditions on $\tilde{z}_0$ and $\hat{B}^\pm_0$.

\end{appendices}


\bibliography{sn-article}

\end{document}